\documentclass[aoas,MSNbibl,nameyear,dvips]{arximspdf}
\usepackage{graphicx}
\usepackage{url,breakurl}
\doi{10.1214/14-AOAS720} %kopijuoti is PTS
\volume{8}
\issue{3}
\pubyear{2014}
\firstpage{1516}
\lastpage{1537}
\docsubty{FLA}

\makeatletter
\def\mathds{\mathbb}
\newcommand{\DIC}{\mathrm{DIC}}
\newcommand{\PPLP}{\mathrm{PPLP}}
\newcommand{\rep}{\mathrm{rep}}
\newcommand{\bolm}{\mathbf{m}}
\newcommand{\bgamma}{\bolds{\gamma}}
\newcommand{\btheta}{\bolds{\theta}}
\newcommand{\bPsi}{\bolds{\Psi}}
\makeatother

\begin{document}
\begin{frontmatter}

\title{Bayesian modeling of bacterial growth for multiple
populations\thanksref{T1}}
\runtitle{Bayesian modeling of bacterial growth}

\begin{aug}
\author[A]{\fnms{A. Paula}~\snm{Palacios}\ead[label=e1]{ana.palacios@plymouth.ac.uk}\thanksref{M1}},
\author[B]{\fnms{J. Miguel}~\snm{Mar\'{\i}n}\ead[label=e2]{juanmiguel.marin@uc3m.es}\thanksref{M2}},
\author[C]{\fnms{Emiliano J.}~\snm{Quinto}\ead[label=e3]{equinto@ped.uva.es}\thanksref{M3}}
\and
\author[B]{\fnms{Michael P.}~\snm{Wiper}\corref{}\ead[label=e4]{michael.wiper@uc3m.es}\thanksref{M2}}
\runauthor{Palacios, Mar\'{\i}n, Quinto and Wiper}
\affiliation{Plymouth University\thanksmark{M1},
Universidad Carlos III de Madrid\thanksmark{M2} and\break 
Universidad de Valladolid\thanksmark{M3}}
\address[A]{A. P. Palacios\\
School of Computing and Mathematics\\
Plymouth University\\
2 Kirkby Place, Drake Circus\\
PL4 8AA, Plymouth\\
United Kingdom\\
\printead{e1}} %adresu isvedimo komanda gale!
\address[B]{J. M. Mar\'{\i}n\\
M. P. Wiper\\
Departamento de Estad\'istica\\
Universidad Carlos III de Madrid\\
Calle Madrid, 126\\
28903 Getafe\\
Spain\\
\printead{e2}\\
\phantom{E-mail: }\printead*{e4}}
\address[C]{E. J. Quinto\\
Facultad de Medicina\\
Universidad de Valladolid\\
Palacio de Santa Cruz\\
Plaza de Sta. Cruz 8\\
47002 Valladolid\\
Spain\\
\printead{e3}}
\end{aug}
\thankstext{T1}{Supported by the Spanish Ministries of Education
Culture and Sports and Economy and Competitiveness.}

% HISTORY:
\received{\smonth{6} \syear{2012}}
\revised{\smonth{8} \syear{2013}}

% ABSTRACT
%
\begin{abstract}
Bacterial growth models are commonly used for the prediction of
microbial safety and the shelf life of perishable foods. Growth is
affected by several environmental factors such as temperature, acidity
level and salt concentration. In this study, we develop two models to
describe bacterial growth for multiple populations under both equal and
different environmental conditions. First, a semi-parametric model
based on the Gompertz equation is proposed. Assuming that the
parameters of the Gompertz equation may vary in relation to the running
conditions under which the experiment is performed, we use feedforward
neural networks to model the influence of these environmental factors
on the growth parameters. Second, we propose a more general model which
does not assume any underlying parametric form for the growth function.
Thus, we consider a neural network as a primary growth model which
includes the influencing environmental factors as inputs to the
network. One of the main disadvantages of neural networks models is
that they are often very difficult to tune, which complicates fitting
procedures. Here, we show that a simple Bayesian approach to fitting
these models can be implemented via the software package \texttt{WinBugs}.
Our approach is illustrated using real experimental \textit{Listeria monocytogenes} growth data.
\end{abstract}

% KEYWORDS
\end{frontmatter}

%s1 #&#
\section{Introduction}\label{sec1}

The predictability of bacterial growth is of major interest due to the
influence of bacteria on food safety and health. The evolution of
microorganisms in food products can spoil the products or even cause
pathogenic effects. Foods are ecosystems composed of the environment
and the organisms that live in it. The food environment is composed of
intrinsic factors inherent to the food (pH, water activity, nutrients)
and extrinsic factors external to it (temperature, gaseous environment,
bacteria). The interactions between the chemical, physical and
structural aspects of a niche and the composition of its specific
microbial population emphasize the dynamic complexity of food
ecosystems [\citet{ICMSF1980}]. Food may contain multiple
microenvironments and can be heterogeneous on a micrometer scale [\citet{montville2005food}]. 
Products in the modern food supply are often
preserved by multiple hurdles that control microbial growth, increase
food safety and extend product shelf life [\citet{leistner2000basic}].
Salt, high- or low-temperature processing and storage, pH, redox
potential and other additives are examples of hurdles that can be used
for preservation [\citet{IOM2010}]. The influence of~pH on bacterial
gene expression is a relatively new area [\citet{montville2005food}].
The expression of genes governing proton transport, amino acid
degradation, adaptation to acidic or basic conditions, and even
virulence can be regulated by the external~pH. The influence of
temperature on microbial growth is very important, both in growth rate
and in gene expression. Cells grown at different temperatures express
different genes (governing from motility to virulence) and are
physiologically different [\citet{MontvilleMatthews2001}]. Salt is
effective as a preservative because it reduces the water activity of
foods (i.e., the amount of unbound water available for microbial growth
and chemical reactions) by the ability of sodium and chloride ions to
associate with water molecules [\citet{fennema1996introduction}; \citet{potter1998food}; \citet{IOM2010}]. Adding salt to foods can also cause
osmotic shock in bacteria cells, limit the oxygen solubility, interfere
with cellular enzymes, or force cells to expend energy to exclude
sodium ions from the cell [\citet{davidson2001}; \citet{shelef2005};
\citet{IOM2010}]. \textit{L.~monocytogenes} is able to grow over a wide range of
temperatures ($-$0.4 to 45$^\circ$C), pH values (4.39 to 9.4) and
osmotic pressures (NaCl concentrations up to 10\%). It is also
facultatively anaerobic [\citet{montville2005food}]. Summarizing, all
these factors can be manipulated to preserve food due to their
influence on the microbial growth. However, even when it is well known
that these factors affect bacterial growth, the kind of effects and the
interactions of the factors are still unclear and need more research.
Accurate models which describe the bacterial growth and the effect of
environmental factors are very important to prevent diseases by
determining the shelf life of perishable foods or by predicting the
behavior of foodborne pathogens.

Starting from \citet{gompertz1825on}, various parametric growth models
which describe the evolution of the population size directly as a
function of time---called primary models---have been developed; see, for
example, \citet{mckellar20042} for a good comparison. These models
perform well in describing the evolution of bacterial density under
fixed experimental conditions. Nevertheless, as described before,
bacterial growth is strongly affected by environmental conditions such
as temperature, acidity or salinity of the environment and, therefore,
when multiple bacterial populations are analyzed, it is important to
account for these effects in growth curve modeling.

In predictive microbiology, models that describe the effect of
environmental conditions on the growth parameters are called secondary
models; see, for example, \citet{rossdalgaard2004}. For example, the
square-root model of \citet{ratkowsky1982relationship} was developed to
describe the effect of suboptimal temperature on growth rates of
microorganisms. This initial approach was later extended to include
other factors such as level of acidity, water activity and salt
concentration in additive or multiplicative models; see, for example,
\citet{mcmeekin1987model}, \citet{miles1997development}, \citeauthor{wijtzes1995modelling} (\citeyear{wijtzes1995modelling,wijtzes2001development}). The most
common secondary models are polynomial models [see, e.g., \citet{mcclure1993predictive}], which allow any of the environmental factors
and their interactions to be taken into account but include many
parameters without biological interpretation. Another important model
class is the cardinal parameter models [see \citet{rosso1995convenient},
\citet{augustin2000modelling} and \citet{pouillot2003estimation}], which
assume that the effect of
environmental factors is multiplicative.

A disadvantage of these models is that they assume simple parametric
forms for the effects of the different environmental factors. Therefore,
more recently, there has been interest in modeling bacteria growth
curves using nonparametric approaches such as artificial neural
networks; see, for example, \citet{hajmeer1997computational}, \citet{geeraerd1998application} and \citet{garcia2002improving}. One
advantage of neural networks is their capability to describe very
complex nonlinear relationships without imposing any structure on the
relationship between the interacting effects. Furthermore, using a
suitable (logistic) basis function which is of a similar shape to
typical bacterial growth curves, neural networks can capture these
curves without the necessity of using large numbers of nodes.

To achieve the general objective of a high level of protection of human
health, food law shall be based on risk analysis [\citet{FAO1995};
\citet{NACMCF1997}; \citet{CEC2002}]. Quantitative microbial risk
assessment\break (QMRA) is the scientific evaluation of the known
or potential adverse health effects resulting from human exposure to
foodborne microbiological hazards. The objective of a QMRA is to derive
a mathematical statement, based on the probability of certain events,
of the chance of adverse health consequences resulting from exposure to
a microbiological agent capable of causing harm [\citet{FAO1995}; \citet{CAC1996}; \citet{NACMCF1997}].
%The microorganism studied and the conditions applied are of great
%interest in food industry and closely related to the efficiency of
%sanitation measures used in food-processing plants.

In this paper, we shall develop two approaches which are applicable to
growth curve estimation for bacterial populations under different
environmental conditions. The first model is based on the Gompertz
function where the dependence of the growth parameters on the
environmental factors is modeled by a neural network. Second, we shall
consider a direct nonparametric approach based on the use of neural
networks as a primary growth model. An important feature of our
approaches is that in cases where we observe bacterial growth in
various colonies under possible different environmental conditions, we
use hierarchical modeling to improve estimation of any single growth
curve by incorporating information from the various different bacterial
populations. Although hierarchical analysis of parametric bacteria
growth models has been undertaken [see, e.g., \citet{pouillot2003estimation}], to the best of our knowledge, hierarchical
analysis has not been combined with nonparametric approaches previously
in this context.

In most empirical work the fitting of any secondary models is carried
out in two steps. First, a primary growth model is fitted to estimate
the growth parameters\vadjust{\goodbreak} and, second, a secondary model is fitted
conditional on the estimated parameters to estimate the controlling
factors. One problem with this strategy is that the estimated
uncertainty of the first stage is not taken into account in the second
stage and, therefore, a poor fit at the first stage could produce
inaccurate estimations at the second stage. Second, most work in
fitting such models has used classical statistical techniques such as
least squares, which, as noted in \citet{pouillot2003estimation}, may
also underestimate uncertainty.

To overcome these problems, inference for our models is undertaken\break
throughout using a Bayesian approach. In the case of the parametric
primary model and neural network secondary model, the use of this
approach avoids the problems inherent in the two-stage inference
outlined previously. Furthermore, our Bayesian approach permits the
prediction of unobserved growth curves and of growth curve values at
future time periods.
% From a food safety point of view, the QMRA generates outputs
%(distributions) used to build
% a risk analysis model applicable to food microbiology. Usually that
%model is obtained after
% a series of Monte Carlo simulations with spreadsheets software (Vose,
%2000).
To our knowledge, neural networks techniques have not been used either
in food risk analysis or with the objective of a QMRA procedure in
mind. We have built a neural network risk model with direct application
in food industry and using very well-known noncommercial software in
the context of Bayesian analysis, because, although previously the
implementation of Bayesian inference for neural networks models has
required the use of complicated sampling algorithms [see, e.g., \citet{lee2004bayesian}], here, we show that inference can be carried out via
the use of the well-known \texttt{WinBugs} software through the
\texttt{R2WinBugs} interface.

The present work covers different issues related to bacterial dynamics:
(i) the use of the hurdle technology with different combinations of
temperatures, pH values and percentage of NaCl with great importance in
ready-to-eat foods safety conditions and in food handling as part of
the foodservice industry; (ii) the use of NN to model the selected
combinations of hurdles because of its absence of imposed restrictions
(i.e., a new approach to the variability of the bacterial behavior
under different environmental conditions and its application to QMRA);
(iii) predictions of new data (interpolate) from the experimental
growth curves obtained in the laboratory (i.e., to obtain proper new
data avoiding the time-consuming and expensive assays carried out in
the laboratory); and (iv) the study of the behavior of \textit{Listeria}
for its application to ready-to-eat foods under the legal requirement
of 100 CFU (colony-forming units)/g or ml established by the EU
Regulation 2073/2005 [\citet{CEC2005}] and the QMRA procedures widely
applied in food industry.

We begin in Section~\ref{sec2} with a brief introduction to neural networks. In
Section~\ref{sec3} we propose two alternative models for bacterial growth curves
that include environmental conditions as influencing factors modeled by
neural networks. In Section~\ref{sec4} we show how to undertake Bayesian
inference for these models and then, in Section~\ref{sec5}, we illustrate the
models with an application to a database of \textit{Listeria
monocytogenes} growth curves generated under various experimental
conditions. Finally, in Section~\ref{sec6}, we present our conclusions and some
possible extensions of our approach.

%s2 #&#
\section{Feedforward neural networks}\label{sec2}

\begin{figure}

\includegraphics{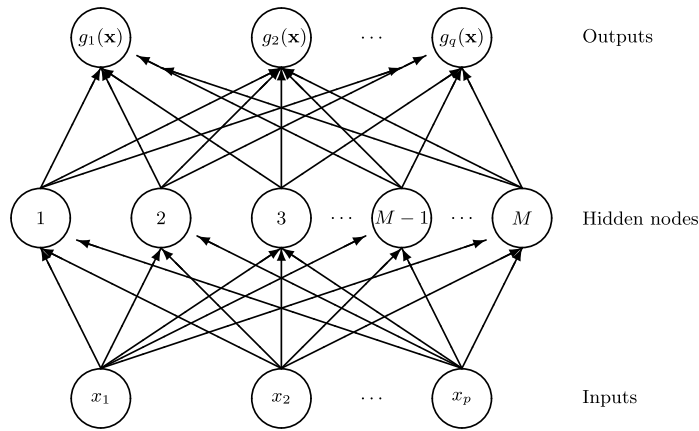}

%%
%%
%(0,0)[l]{Outputs}}
%(0,0)[c]{$\ldots$}}
%%
%%
\caption{Neural network representation.}\label{nnfig}
\end{figure}

In many situations it is assumed that there are $q$ dependent
variables, $(Y_1, \ldots, Y_q)=\mathbf Y$, and they can be modeled as an
approximate linear or polynomial function of a set of explanatory
variables, $(x_1, \ldots, x_p)= \mathbf x$, via, for example, multivariate
regression. However, such a relationship may not always be
appropriate and a more general functional relation between the
dependent and independent variables must be assumed, say,
\[
E[\mathbf{Y}|{\mathbf{x}}] = \mathbf{g}({\mathbf{x}}),
\]
where the functional form, $(g_1,\ldots,g_q)=\mathbf{g}\dvtx \mathds
{R}^p\rightarrow\mathds{R}^q$, is unknown. One of the most popular
methods of modeling the function $g$ is via neural networks; see, for
example, \citet{stern1996neural}. In particular, a feedforward neural
network takes a set of inputs $\mathbf x$ and from them computes the vector
of output values as follows:
%
%
%e1 #&#
\begin{equation}
\label{NN} \mathbf{g}(\mathbf{x})= B \cdot\bPsi^T\bigl(
\mathbf{x}^T\Gamma\bigr),
\end{equation}
where $B$ is a $q \times M$ matrix with $q\in\mathds{N}$ the number
of output variables and $M \in\mathds{N}$ the number of nodes and
$\Gamma$ is a $p \times M$ matrix with $p\in\mathds{N}$ being the
number of explicative variables. The element $\gamma_{rk} \in\mathds
{R}$ is the weight of the connection from input~$r$ to hidden unit $k$
and the element $\beta_{sk} \in\mathds{R}$ is the weight connection
from hidden unit $k$ to output unit $s$. Finally, $\bPsi(a_1, \ldots,
a_M)=(\Psi(a_1),\ldots,\Psi(a_M))$, where $\Psi$ is a sigmoidal
function such as the logistic function
%
%
%e2 #&#
\begin{equation}
\label{logistic} \Psi(x) = \frac{\exp(x)}{1+ \exp(x)},
\end{equation}
which we will use here, as typically bacterial growth curves have an
approximately sigmoidal form. Equations (\ref{NN}) and (\ref
{logistic}) define a feedforward neural network with logistic
activation function, $p$ explanatory variables (inputs), one hidden
layer with $M$ nodes and $q$ dependent variables (outputs) that can be
illustrated as in Figure~\ref{nnfig}.

%
%
%f1 #&#

Note that each output combines the node values in a different way. For
practical fitting of neural networks models, it is typically assumed
that the input variables are all defined to have a similar finite
range, for example, $[0,1]$. From now on, we shall assume this throughout.

%s3 #&#
\section{Neural network-based growth curve models}\label{sec3}

\begin{figure}

\includegraphics{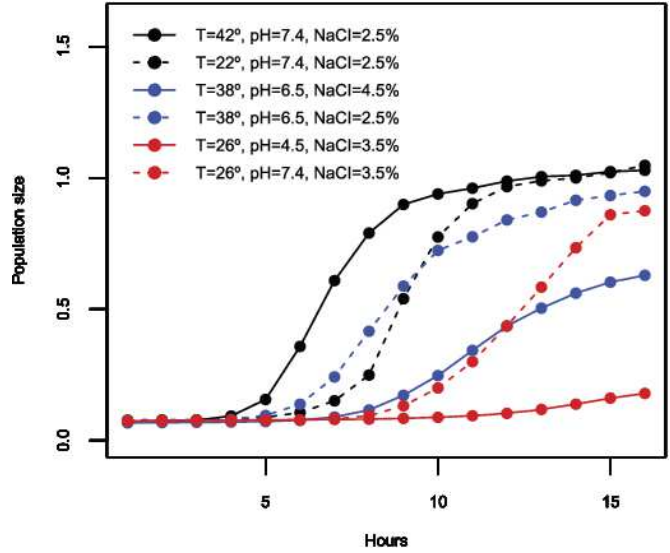}

\caption{Bacterial growth under different environmental conditions.}\label{figfactors}
\end{figure}

Bacterial growth is very influenced by environmental factors. For
example, bacteria grow in a wide range of temperatures, but in higher
temperatures bacterial growth increases and in lower temperatures it
decreases. In a similar way, changes on the level of acidity or
salinity affect the growth of bacteria. The grade and the direction of
the effect depend on the strain of bacteria and also on the level of
the other factors. Figure~\ref{figfactors} shows the different
behaviors of \textit{Listeria} growth under different environmental conditions.

%
%
%f2 #&#

To account for these effects, we develop growth curve models based on
the use of neural networks.
% to explain the functional relationship of growth to given
%environmental factors.

%s3.1 #&#
\subsection{A neural network-based Gompertz model}\label{sec3.1}

The bacterial growth process is typically characterized by three
distinct phases, that is, the lag stage that reflects the adaptation of
cells inoculated in a new medium; the exponential stage that represents
the bacterial growth by binary fission; and, finally, the stationary
stage which describes the decay of the growth rate as a consequence of
nutrient depletion and accumulation of waste which is followed by death
or decline of the population. Sigmoidal functions which account for
these three phases have been typically used to model microbial growth;
see, for example, \citet{skinner1994mathematical}. In particular, the
Gompertz equation is a well-known model for bacterial growth over time
and it has been used extensively by researchers to fit a wide variety
of growth curves from different microorganisms; see, for example, \citet{ross1994predictive} and \citet{mckellar20042}.

Here we consider a reparameterized Gompertz equation proposed by \citet{zwietering1990modeling}. Let $N_t$ represent the population
concentration of bacteria cultivated in a Petri dish experiment at time
$t \ge0$. Then the Gompertz equation is
%
%
%e3 #&#
%e4 #&#
\begin{eqnarray}\label{Gomp}
E[N_t|N_{0},D,\mu,\lambda]  =  g(t,N_0,D,
\mu,\lambda)
\nonumber\\[-8pt]\\[-8pt]
\eqntext{\displaystyle\mbox{where }
g(t,N_0,D,\mu,\lambda)  =  N_0 + D \exp\biggl( -\exp
\biggl(1+ \frac{\mu e (\lambda- t)}{D} \biggr) \biggr),}
\end{eqnarray}
where $e$ is Euler's number, $N_0$ is the initial bacterial density,
$D$ is the difference between the maximum bacterial density, $\mu$ is
the maximum growth rate and $\lambda$ is the time lag.

The primary growth model described in (\ref{Gomp}) does not allow for
the case where we wish to study bacterial populations under a variety
of controlled environmental conditions. Thus, suppose that we observe
the growth of $I$ bacterial populations under similar initial
conditions and that we have $J$ different environments determined by
temperature, level of acidity (pH) and salt concentration (NaCl). Under
fixed environmental conditions, it may be reasonable to assume that all
replications have the same growth curve parameters. However, growth
rates will vary under different conditions and, therefore, assuming a
Gompertz model, we propose the use of neural networks to reflect the
parameter dependence on the environmental factors. If $N_{tij}$ is the
concentration in population $i$ under environmental conditions $j$ at
time $t$, the Gompertz function is
%
%
%e5 #&#
\begin{equation}
\label{GompNN} E[N_{tij}|N_{0j},D_j,
\mu_j,\lambda_j] = g(t_{ij},N_{0j},D_j,
\mu_j,\lambda_j),
\end{equation}
where $g(\cdot)$ is as in (\ref{Gomp}),
for $i=1,\ldots,I$ and $j=1, \ldots, J$. Now, we model the growth
parameters $\mu$, $\lambda$ and $D$ as a function of the temperature,
the level of acidity and the salt concentration by a feedforward neural
network, that is,
%
%
%e6 #&#
\begin{equation}
\label{GompNN2} \btheta_s = \sum_{k=1}^M
\beta_{sk} \cdot\Psi\bigl(\mathbf{x}'\bgamma
_k\bigr)\qquad\mbox{for $s=1,2,3$,}
\end{equation}
where $\btheta_s$ stands for the parameters $D, \mu, \lambda$ and
$\mathbf{x}= (T, \mbox{pH}, \mbox{NaCl})$ is the vector of explanatory variables and
$\Psi$ is the logistic function. Note that this network does not
include an intercept term. In our practical experiments we have found
that the addition of an intercept produces no significant differences
to typical curve fits.
The model defined in this section by expressions (\ref{GompNN}) and
(\ref{GompNN2}) will be referred to as the GNN model.

%s3.2 #&#
\subsection{A hierarchical neural network model}\label{sec3.2}

Here, we generalize the previous model to a new one which does not
assume any underlying parametric growth function. Instead, we propose a
neural network as a primary model. The output of the network is the
instantaneous reproduction rate per member of the population and the
inputs are the current population size and the experimental conditions.
Formally, we can write the model as
%
%
%e7 #&#
%e8 #&#
\begin{eqnarray}\label{NN2}
&& E[N_{tij}|N_{(t-1)ij},f_{j},T_{j},\mbox{pH}_{j},\mbox{NaCl}_{j}]
\nonumber\\[-8pt]\\[-8pt]
&&\qquad =
N_{(t-1)ij}+N_{(t-1)ij}f_{j}(N_{(t-1)ij},T_{j},\mbox{pH}_{j},\mbox{NaCl}_{j}),\nonumber
\\
&& f_{j}(N_{(t-1)ij},T_{j},\mbox{pH}_{j},\mbox{NaCl}_{j})\nonumber
\\
&&\qquad =
\sum_{k=1}^{M}\beta_{jk}\bigl(
\Psi(\gamma_{1k}N_{(t-1)ij}+\gamma_{2k}T_{j}+
\gamma_{3k}\mbox{pH}_{j}+\gamma_{4k}\mbox{NaCl}_{j})
\\
&&\hspace*{119pt}{}-\Psi(\gamma_{2k}T_{j}+\gamma_{3k}\mbox{pH}_{j}+
\gamma_{4k}\mbox{NaCl}_{j})\bigr),\nonumber
\end{eqnarray}
for $i=1,\ldots,I$ and $j=1,\ldots,J$, and $f_{j}(\cdot)$ is the
growth rate for populations with environmental condition $j$. As
previously, we could consider adding an intercept term to the network.
However, for the given model, given the addition of an error term as
defined in the following subsection, when $N_{(t-1)ij} = 0$, then
$N_{tij} = 0$, so that once the population has died out, then it
remains extinct. Including an intercept would mean that this desirable
property is lost. The model defined in this section by (\ref{NN2})
will be referred to as the NN model.

%
%
%f3 #&#
\begin{figure}%[hbt]

\includegraphics{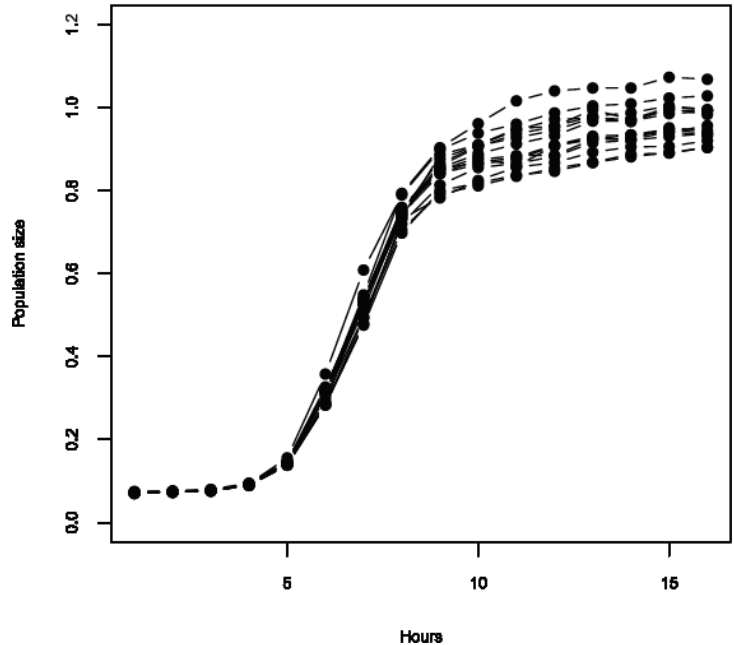}

\caption{15 replications of bacterial growth under $T=42^\circ$C, pH${}=7.4$ and NaCl${}={}$2.5\%.}\label{figcurves}
\end{figure}

%s3.3 #&#
\subsection{Error modeling}\label{sec3.3}

In the previous subsections two approaches to modeling the expected
population density have been provided. These models are completed by
including an error term. Thus, in the case of the full neural network
model, we assume that
%
%
%e9 #&#
\begin{equation}
N_{tij} = N_{(t-1)ij} + N_{(t-1)ij}f_{j}(N_{(t-1)ij},T_j,\mbox{pH}_j,\mbox{NaCl}_j)
+ \varepsilon_{tij},
\end{equation}
where we assume that the error term is%
%
%e10 #&#
\begin{equation}
\varepsilon_{tij}|N_{(t-1)ij},\sigma,v \sim{\mathcal N} \bigl(0,
\sigma^2 N_{t-1}^v \bigr),
\end{equation}
where $\sigma^2 \ge0$ and $v=0.5$ so that the possibility that the
error variance increases with population density is allowed for.
Figure~\ref{figcurves}
illustrates different bacterial growth curves from Petri dish
experiments under the same conditions. It can be seen that the curves
are closer together initially when the population density is lower and
diverge over time as the population density grows, which suggests that
a model of this type is reasonable. Our empirical experiments suggest
that the value of $v = 0.5$ is appropriate here, although, clearly, a
prior distribution for $v$ could be considered. Following the same idea
of increasing error variance, we assume for the GNN model that the
error term is%
%
%e11 #&#
\begin{equation}
\varepsilon_{tij}|g{t_{ij}},\sigma,v \sim{\mathcal N} \bigl(0,
\sigma^2 g(t _{ij})^v \bigr),
\end{equation}
where $g(\cdot)$ is the Gompertz function evaluated at the current time point.

%s4 #&#
\section{Bayesian inference for the neural network models}\label{sec4}

Given a set of observed inputs and outputs from a neural network, say,
$D=(x_1,y_1), \dots,\break (x_N,y_N)$, inference can be carried out using a
variety of approaches; see, for example, \citet{neal1996bayesian} and
\citet{fine1999feedforward} for reviews. Here, we shall consider a
Bayesian approach.
To implement such an approach, we must first define suitable prior
distributions for the neural network parameters $\beta$ and $\gamma$
and for the uncertainty. First, we suppose little prior knowledge
concerning the variance and, hence, we propose a vague inverse-gamma
prior distribution for it, $\sigma^{-2} \sim G(a/2,b/2)$. In neural
network models, it is common to use relative uninformative prior
distributions due to the scarcity of prior information about the
parameters. For simplicity, we choose normal and gamma distributions
with a hierarchical structure, that is,
\begin{eqnarray*}
% p & \sim& {\mathcal U}[0,1] \\
\beta_{ik}|m_{i \beta},
\sigma^2_{\beta} & \sim& {\mathcal N} \bigl(m_{i \beta},
\sigma^2_{\beta} \bigr),
\\
\bgamma_k|m_{\gamma
},\sigma^2_{\gamma} &
\sim& {\mathcal N} \bigl(\bolm_{\gamma},\sigma^2_{\gamma}I
\bigr),
\end{eqnarray*}
where the subscript $i$ in the GNN model accounts for the growth
parameters and in the NN model for the groups defined by the
environmental conditions. The Bayesian approach is completed by vague,
but proper prior distributions for the remaining hyperparameters as follows:
\begin{eqnarray*}
m_{i\beta}|\sigma^2_{\beta} & \sim& {\mathcal N}
\biggl(m_{0\beta}, \frac{\sigma^2_{\beta}}{c_{\beta}} \biggr),
\\
m_{0\beta} |\sigma^2_{\beta} & \sim& {\mathcal N}
\biggl(0,\frac{\sigma
^2_{\beta}}{e_{\beta}} \biggr),
\\
\frac{1}{\sigma^2_{\beta}} & \sim& {\mathcal G} \biggl( \frac{d_{\beta
1}}{2},
\frac{d_{\beta2}}{2} \biggr),
\\
\bolm_{\gamma}|\sigma^2_{\gamma} & \sim& {\mathcal N}
\biggl(\mathbf{0}, \frac{\sigma^2_{\gamma}}{c_{\gamma}} I \biggr),
\\
\frac{1}{\sigma^2_{\gamma}} & \sim& {\mathcal G} \biggl( \frac
{d_{\gamma1}}{2},
\frac{d_{\gamma2}}{2} \biggr),
\end{eqnarray*}
where $c_{\beta},e_{\beta}$, $d_{\beta1}$, $d_{\beta2}$, $c_{\gamma
}$, $d_{\gamma1}$ and $d_{\gamma2}$ are assumed known and fixed.
Similar hierarchical prior distributions are typically used in Bayesian
inference for neural network models; see, for example, \citet{lavine1992bayesian}, \citet{muller1998iba} and \citet{andrieu2001robust}. For alternatives, see, for example, \citet{lee2004bayesian}, \citet{robert1999reparameterisation} and \citet{roeder1997practical}.

Usually, we will have good prior knowledge about the average initial
population density,
$m_0 = E[N0_{i}|m_0,s_0]$, and the variance, $s_0$, as typically Petri
dishes are seeded with very similar quantities of bacteria close to a
known theoretical level, so we shall typically assume that these are
known. Otherwise, a simple noninformative prior distribution
$f(m_0,t_0) \propto1/t_0$, where $t_0 = 1/s^2_0$ can be used when,
immediately, we have that given the observed set of initial densities,
$\mathbf{N0} = (N0_1,\ldots,N0_I)$,
\begin{eqnarray*}
m_0|\mathbf{N0},s_0 & \sim& {\mathcal N} \biggl(\overline{N0},\frac{s^2_0}{I} \biggr),
\\
s^2_0|\mathbf{N0} & \sim& {\mathcal IG} \Biggl(I-1, \sum
_{i=1}^I (N0_i-
\overline{N0})^2 \Biggr),
\end{eqnarray*}
where $\overline{N0} = \frac{1}{I} \sum_{i=1}^I N0_i$ is the average
initial density and $\mathcal IG $ means inverse gamma.

Given the above prior structure, a closed form for the posterior
parameter distributions is not available. However, Markov Chain Monte
Carlo (MCMC) techniques can be employed to allow us to generate an
approximate Monte Carlo \mbox{sample} from the posterior parameter
distributions; see, for example, \citet{gilks1996markov} for a full
review. Various different MCMC algorithms have been proposed in the
neural networks literature, but in general the efficiency of such
samplers depends on the model; see, for example, \citet{lee2004bayesian}.

As an alternative, here, we propose using the generic MCMC sampler,
\mbox{\texttt{WinBugs}}, as developed by \citet{spiegelhalter1999winbugs},
which is appropriate for hierarchical modeling situations, programmed
in combination with~\texttt{R}, via \texttt{R2WinBugs}.

\begin{figure}%[hbtp]

\includegraphics{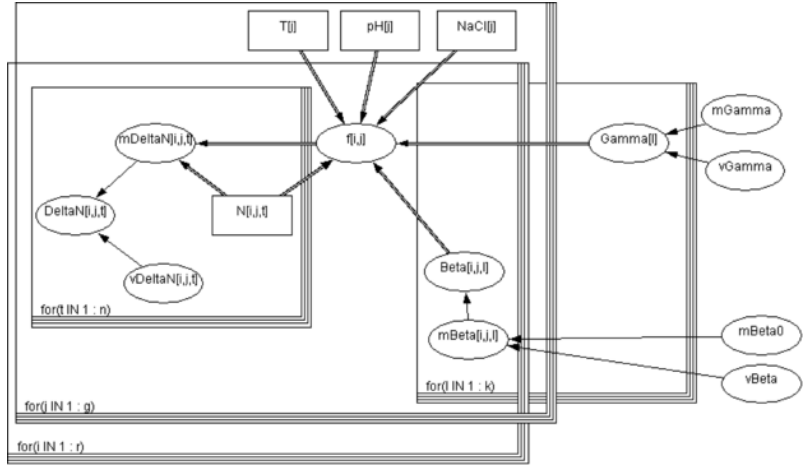}

\caption{Dependence structure of the NN model.}\label{doodle}
\end{figure}

Figure~\ref{doodle} illustrates the dependence structure of the NN
model in \texttt{WinBugs} style (although code cannot be constructed
directly from this diagram). In the figure, random and logical nodes
are represented by ellipses and fixed nodes (independent variables) are
represented by rectangles. The arrows represent dependence
relationships, with the single arrows showing stochastic dependence and
the double arrows representing logical dependence. For more details see
\url{http://www2.mrc-bsu.cam.ac.uk/bugs/winbugs/contents.shtml}.
%
%
%f4 #&#

As \texttt{WinBugs} is a generic approach to MCMC sampling, it is
important to check on the convergence of the sampler. Various tools can
be used to check the convergence. In particular, as well as standard
graphical techniques such as looking at the trace, the evolution of the
mean and the autocorrelations of the sampled output, we also use formal
diagnostic techniques such as the modified Gelman--Rubin statistic, as
in \citet{brooks1998general}.

Note finally that the codes for running both models are available in
the supplemental materials [\citeauthor{supplementA} (\citeyear{supplementA,supplementB})].

%s4.1 #&#
\subsection{Model selection}\label{sec4.1}

Thus far, inference is conditional on the number of hidden nodes, $M$,
being known. Various approaches to estimating $M$ may be considered.
One possibility is to treat $M$ as a variable and, given a prior
distribution for $M$, use variable-dimensional MCMC approaches to carry
out inference; see, for example, \citet{muller1998iba} or \citet{neal1996bayesian}. Another approach which we shall use in this article
is to use an appropriate model selection technique to choose the value
of $M$.

A number of criteria have been proposed for model selection in Bayesian
inference. A standard Bayesian selection criterion which is
particularly appropriate when inference is carried out using MCMC
methods is the deviance information criterion (DIC), as proposed in
\citet{spiegelhalter2002bayesian}. However, in the context of neural
networks, the possible lack of identifiability of the model or
multimodality of the posterior densities make this criterium unstable.
Many variants of the DIC have also been considered and, here, we prefer
to apply the $\DIC_3$ criterion of \citet{celeux2006deviance}. For a
model ${\mathcal M}$ with parameters $\btheta$ and observed data $\mathbf{y}$, the $\DIC_3$ is defined as follows:
\[
\DIC_3 = -4E_{\btheta}\bigl[\log f(\mathbf{y}|\btheta)|\mathbf{y}
\bigr] + 2\log\prod_{i=1}^n
E_{\btheta}\bigl[f(y_i|\btheta,\mathbf{y})\bigr].
\]
In \citet{celeux2006deviance} this criterion is recommended in the
context of latent variable models. Furthermore, \citet{watanabe2010}
recommends the use of this criterion in the case of singular models
such as neural networks.

An alternative approach which we also consider when comparing different
models is the posterior predictive loss performance (PPLP) proposed by
\citet{gelfand1998model}. Based on the posterior predictive
distribution, this criterion consists in defining a weight loss
function which penalizes actions for departure from the corresponding
observed value as well as for departure from what we expect the
replication to be. In this way, the approach is a compromise between
the two types of departures: fit and smoothness. For squared error
loss, the criterion becomes
\[
\PPLP=\frac{k}{k+1} \sum_{i=1}^n
(m_i - y_i)^2+\sum
_{i=1}^n s_i^2,
\]
where $m_i=E[y^{\rep}_i|\mathbf{y}]$ and $s_i^2=\operatorname{Var}[y^{\rep}_i|\mathbf{y}]$
are, respectively, the mean and the variance of the predictive
distribution of $y^{\rep}_i$ given the observed data ${\mathbf y}$ and $k$
is the weight we assign to departures from the observed data. The first
term of the $\PPLP$ is a plain goodness-of-fit term and the second term
penalizes complexity and rewards parsimony.

%s5 #&#
\section{Application: \textit{Listeria monocytogenes}}\label{sec5} \label{listeria}

In this section we analyze a data set taken from Petri dish experiments
of one of the authors (EQ) and consisting of measures of the
concentrations of \emph{Listeria monocytogenes} bacteria in a Petri
dish under several experimental conditions. A strain of \textit{Listeria monocytogenes} previously isolated from poultry meat was
provided by the Department of Animal and Food Sciences, School of
Veterinary Medicine, Autonomous University of Barcelona, Spain, and
used in the present study. \textit{L.~monocytogenes} growth data was obtained as
reported by \citet{eduardo2011analyzing}. Briefly, an automated method
(SLT 340 ATTC microplate reader, SLT Labinstruments, Austria) for the
measurements of the optical density of a \textit{L.~monocytogenes} culture was
used. Aliquots of the microorganism, previously cultured in nutrient
broth at 31$^\circ$C overnight and serially diluted, were inoculated
into the microplate wells and read at a wavelength of 595 nm every 15
min. Optical density curves of bacterial growth were obtained. At the
same time, aliquots were also spread onto Petri plates with nutrient
agar and cultured at 31$^\circ$C overnight. The environmental factors
taken into account are temperature, level of acidity and salinity.
Temperatures range between 22$^\circ$C and 42$^\circ$C, pH between
4.5 and 7.4 and NaCl between $2.5\%$ and $5.5\%$. There are $96$
different combinations of environmental factors (we call groups) and
for each group there are several replications. The number of
observations per curve varies between 16 and 24, depending on the
curve. We kept for the analysis $74$~groups (excluding the cases with
extreme values of factors which inhibit growth) and chose randomly 10
replications for each one so that the remaining curves could be used
for cross-validation and prediction purposes and to reduce
computational time. The temperatures selected cover the following
situations in food handling: room \mbox{temperature} in northern countries
(22 and 26$^\circ$C); room temperature in warm countries (30 and
34$^\circ$C); and inadequate reheating treatments of ready-to-eat foods
previous to consumption (38 and 42$^\circ$C). The selected pH values
cover most of the range of the pH values tolerated by \textit
{Listeria}. The percentages of NaCl selected are well under the limits
tolerated by \textit{Listeria}, but it is very important to know their
possible effects under a hurdle technology point of view combined with
temperature and pH values. A reduced version of this data set including
six groups under the same environmental conditions as the data
illustrated in Figure~\ref{figfactors} and each with ten replications
is contained in the supplemental materials [\citet{supplementA}].

Using the $\DIC_3$ criterion as outlined earlier, the optimum number of
nodes for both models is 2. Temperature, pH and NaCl as inputs of the
neural networks were previously scaled onto [0.1, 0.9] as recommended
in \citet{valero2007searching}. In the implementation of the GNN model
we keep the hyperparameters $m_{i\beta}$, $\sigma_{\beta}$, $
m_{\gamma}$ and $\sigma_{\gamma}$ fixed at $m_{i\beta}=0$, $\sigma
_{\beta}=10$, $\bolm_{\gamma}=(0,\ldots,0)'$ and $\sigma_{\gamma
}=10$. Regarding the error variance, we choose $a=0.2$ and $b=0.2$. In
the NN model
% we keep the hierarchical structure presented in section \ref{infer}
%for parameters $\beta_{ijk}$ (where $i$ indicates the replication
%number, $j$ the group of environmental conditions and $k$ the node)
%and $\bgamma_{k}$.
the highest level of hyperparameters were set to $c_{\beta}=10$,
$e_{\beta}=10$, $d_{\beta1}=0.1$, $d_{\beta2}=0.01$, $c_{\gamma
}=10$, $d_{\gamma1}$ and $d_{\gamma2}=0.01$.

For both models, we generated chains with random initial values and
200,000 iterations each, including 100,000 iterations of burn-in. To
diminish autocorrelation between the generated values, we also used a
thinning rate of $1000$. Trace plots and autocorrelation functions were
used to check convergence in the predictions and in all cases it was
found that the burn-in period of 100,000 iterations was reasonable.
Furthermore, the Gelman--Rubin statistic was equal or very close to $1$
for predictions, being a good indicator of convergence.

In order to have a benchmark for the comparison of models, we also fit
two different simple models, the independent Gompertz model and the
pooled Gompertz model. The first one implies that each observed curve,
including the replications, is independent and therefore has its own
Gompertz growth parameters. Independent normal prior distributions with
mean zero and variance 100 are assumed for these parameters. In
contrast, the pooled model assumes that the replications under a fixed
set of environmental conditions are samples from a unique underlying
growth curve for that set of conditions. Normal priors are then placed
on the parameters of this growth curve as for the independent model.
For both benchmark models the errors are the same as in the GNN case
with a ${\mathcal G}(0.1,0.1)$ prior distribution for the error variance.

%
%
%t1 #&#
\begin{table}%[hbt]
\tabcolsep=0pt
\tablewidth=200pt
\caption{Model comparison}\label{DIC}
\begin{tabular*}{\tablewidth}{@{\extracolsep{\fill}}@{}lcc@{}}
\hline
\textbf{Model} & \textbf{DIC\tsub{3}} & \textbf{PPLP}\\ %& G & P \\
\hline
Independet Gompertz & $-$19,136& 781\\
Pooled Gompertz & $-$39,420& $211$\\
Gompertz \& NN & $-$40,099 & \phantom{0}$41$ \\%& & \\
Neural Networks & $-$58,492 & \phantom{0}$28$\\
\hline
\end{tabular*}
\end{table}

%
%
%f5 #&#
\begin{figure}[b]

\includegraphics{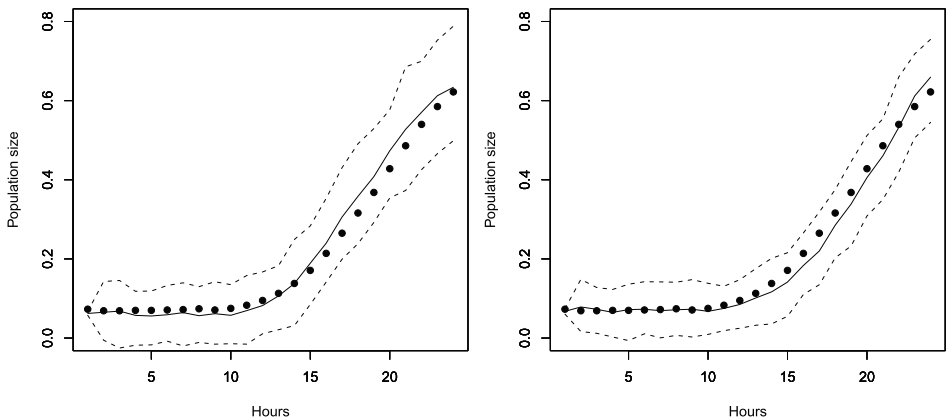}

\caption{Fitting of the GNN model \textup{(left)} and the NN model \textup{(right)}. Points
represent real data, the solid lines represent the posterior means and
the dashed lines represent the 95\% credible interval.}\label{fit}
\end{figure}

The $\DIC_3$ and the $\PPLP$ criteria were computed to compare the
different models under consideration and Table~\ref{DIC} shows the
estimated values for all of these models. As is expected, the pooled
model performs better than the independent one since the assumption of
independence for all the curves is somewhat extreme. Therefore, it
seems reasonable to assume different curves under different
environmental conditions, but under equal conditions we assume a common
curve and this is the approach we choose for the proposed models. But
the problem with this model is that it does not explain the effect of
the environmental factors and it is needed to estimate one model for
each group of conditions. Then, regarding our proposed models which
incorporate the environmental factors as explanatory variables, the
results show that the hierarchical neural network model outperforms the
Gompertz model with neural networks for the parameters. The $\DIC_3$ and
the $\PPLP$ values are lowest for the former model.

Figure~\ref{fit} shows for a particular curve ($T=34^\circ$C,
pH${}=6.5$ and $\mbox{NaCl}=5.5\%$) the fitting of both models. On the left, the
Gompertz model with neural networks explains the dependence of the
growth parameter on the environmental factor and on the right the
fitting of the hierarchical neural network model. The observed values
are represented by points, the estimated growth curves are represented
by the solid line, and the dashed lines represents the $95\%$ credible
interval computed from the posterior distributions. It can be observed
that the fit is good in both cases and the credible intervals included
all the true observations. Nevertheless, note that the NN model
overestimates the lag period. In the remaining curves (replications and
different group conditions), we also found good fits for both models.
Similar results are observed in the fitted plots for all the groups.

%
%
%f6 #&#
\begin{figure}

\includegraphics{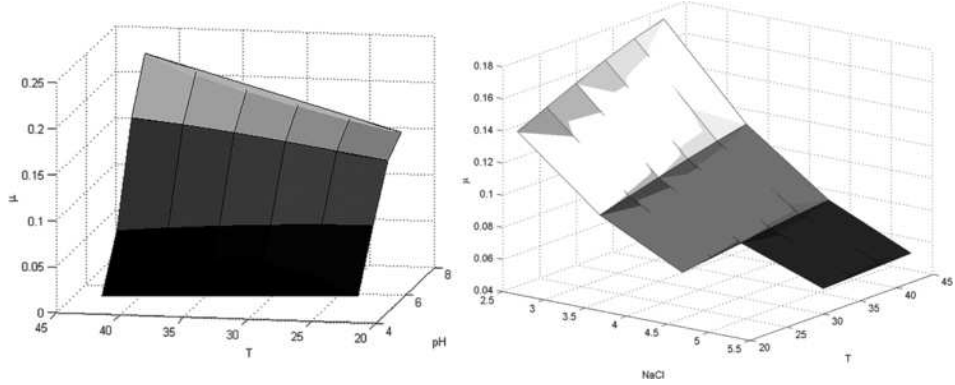}

\caption{Posterior mean of the growth rate parameter $\mu$ for the GNN model. NaCl${}=2.5$\%  (left) and pH${}=6.5$ (right).}
\label{postlag}
\end{figure}

Additionally, with the GNN model we can make predictions of the growth
parameter values for a certain level of environmental factors. Based on
previous works, it should be expected that an increase of temperatures
and a decrease in pH values kills a foodborne pathogen. However,
predictions from our model show an interesting behavior of \textit
{Listeria} under several environmental conditions. The impact of
temperature on growth is not the same when considering different pH
values, changing even the direction of the effect. On the other hand,
the effects seem to be irregular and interacting, which emphasizes the
utility of a neural network model which does not impose a rigid
functional form on the dependencies. To illustrate these effects, we
plot the posterior mean of the growth rate parameter as a function of
the environmental factors (see Figure~\ref{postlag}).

For example, when pH values range between 4.5 and 5.5, an increase in
the temperature values is needed to decelerate the growth rate of
\textit{Listeria}. In contrast, when pH is equal to 6.5 or 7.4, the
temperature must be decreased to diminish the microorganism growth.

Regarding the percentage of NaCl, we found a decrease in the growth
rate when the percentage of NaCl increases. Additionally, the impact
grade of the temperature changes for different values of NaCl. When
NaCl is equal to $5.5\%$ the differences among the temperature effects
are minimal, but those differences increase for lower levels of NaCl.

\begin{figure}%[ht]

\includegraphics{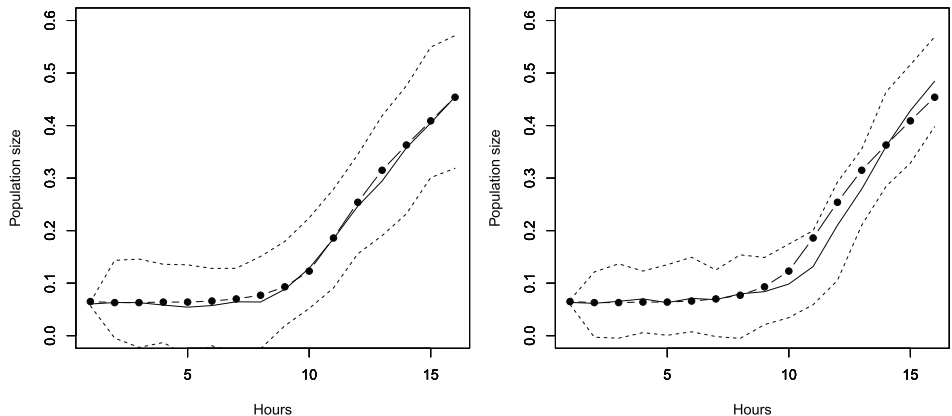}

\caption{One-step-ahead predictions of the GNN model \textup{(left)} and the NN
model \textup{(right)}. Points represent real data, the solid lines represent the
posterior means and the dashed lines represent the 95\% credible interval.}\label{paso}
\end{figure}

Now, we consider predictions of future values of a growth curve and
predictions of a full curve under an unobserved set of environmental
conditions. For the first case, we computed one-step-ahead predictions.
That is, for a particular curve we observe data until observation $t$
and predict the population size at $t+1$. In the next step, we observe
data until $t+1$ and predict the population size at $t+2$ and so on,
until the completion of the predictive curve. Figure~\ref{paso} shows
the one-step-ahead predictive curves for both models for a particular
growth curve ($T=42^\circ$C, pH${}=5.5$ and $\mbox{NaCl}=2.5\%$). In contrast
to the fitting results, the Gompertz model shows a slightly better
performance regarding the mean prediction. The mean square error of the
prediction in the Gompertz model is equal to 0.001, while for the NNs
model it is 0.008. But in the second model higher accuracy is reached,
as can be seen from the narrower credible interval.

%
%
%f7 #&#

\begin{figure}%[ht]

\includegraphics{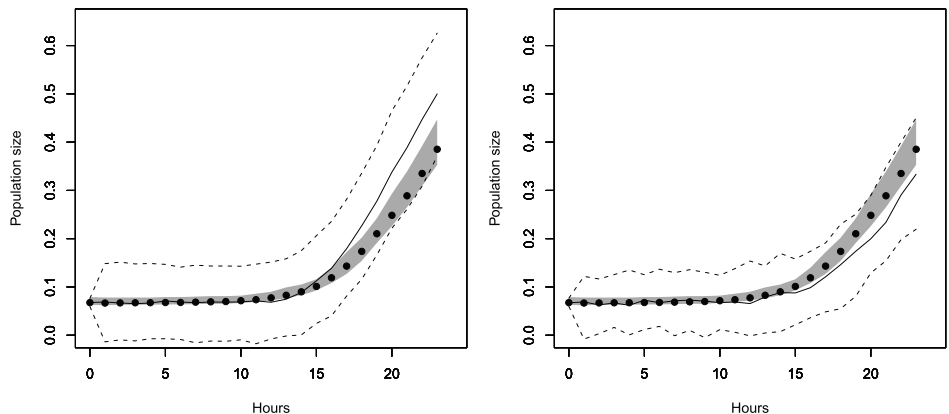}

\caption{Prediction out of sample of the GNN model \textup{(left)} and the NN
model \textup{(right)}. Shadings represent the area where real data of all
replications lie, the solid lines represent the posterior means and the
dashed lines represent the 95\% credible interval.}\label{cross}
\end{figure}

In the context of model checking, several authors, for example, \citet{gelfand1996model} and \citet{vehtari2003expected}, have proposed the
use of cross-validatory predictive densities. Following this approach,
the data set is divided in two subsets $({\mathbf y_1}, {\mathbf y_2})$. The
first subset is used to fit the model and to estimate the posterior
distribution of the parameters, while the second set is used to compute
the cross-validatory predictive density: $f({\mathbf y_1}| {\mathbf y_2})= \int
f({\mathbf y_2}|\btheta)f(\btheta|{\mathbf y_1}) \,d\btheta$. In our case, we
computed the predictive density for one of the groups which was not
used in the model fitting. Given the hierarchical structure of the
models, it is possible to make predictions of a growth curve under an
unobserved set of conditions, due to the knowledge learned from the
other observed group of conditions. To illustrate, we make predictions
for a new unobserved group with $T=26^\circ$C, pH${}=6.5$ and
$\mbox{NaCl}=5.5\%$. Figure~\ref{cross} shows the mean prediction (solide
line) and the $95\%$ credible interval (dashed line) for both models,
GNN on the left and NN on the right. As there are many replications for
this group, we plot only the mean curve and shade the area between the
minimum value and the maximum value observed for each time $t$ among
replications. As an input of the neural network for the NN model we
used the mean curve of the replications.

%
%
%f8 #&#

The out-of-sample predictions of both models are fairly good and
constitute one of the main contributions of this work. Although the
good performance of both models, in the case of the GNN model some
observations lie outside the credible interval---a small shaded area
lies outside the dashed line. Moreover, comparing the mean prediction
with the mean observed curve, the NN model yields more accurate predictions.

%s6 #&#
\section{Conclusions and extensions}\label{sec6}

In this paper we have shown a me\-tho\-do\-lo\-gi\-cal contribution
which can be easily and directly applied for microbiological
researchers. Neural networks were used as a secondary model that
explains the dependence on environmental factors and also as a primary
model which, besides time, includes experimental conditions as
explanatory variables. Inference was carried on in a Bayesian approach
that avoids the problems for doing inference in two steps. Both models
yield accurate estimations and good predictions which show that NNs can
be used to model bacterial growth, describing accurately the complex
interacting effects of environmental factors without imposing any
simplifying assumption.
On the other hand, the modified Gompertz equation was used as the base
model for the first approach we considered, but other parametric
bacteria growth
models such as Baranyi or logistic are equally applicable.

Estimations were implemented in \texttt{WinBugs} via \texttt{R2WinBugs},
showing that \texttt{WinBugs} can be a powerful and flexible tool that is
able to handle very complex models such as neural networks with great
ease. As \citet{mackay1995bayesian} pointed out, the Gibbs sampling
method is not the most efficient of MCMC methods, but there may be
problems of interest where the convenience of this tool outweighs this drawback.

Previous studies have special interest for the food industry. The
conditions inside a food-processing plant (humidity, temperature, food
processing techniques, sanitation procedures, etc.) relate to each
other in a very complex way, creating microenvironments with adequate
conditions for the growth of \textit{Listeria}, such as hard-to-reach
areas (drains, etc.). The use of NNs gives more flexibility, as they do
not impose restrictions to the hurdle technology effects on
microorganisms and can show more freely the variability inherent to any
form of life under different environmental conditions. And it is
necessary to take into account that variability does not only appear in
a laboratory assay, but also and most importantly appears in a food
industry production chain, in a foodservice company, in a food
distributor or at home before the moment of consumption. Following this
reasoning, the application of the NNs to quantitative microbial risk
assessment seems a very useful and realistic tool, reflecting with
fewer restrictions the behavior of foodborne pathogens. This
flexibility in the model has allowed us to get new conclusions,
different to previous studies.

Food safety conditions and food handling are part of the foodservice
industry, and different conditions of temperature, pH and percentage of
NaCl give a new insight in terms of inhibitory effects of those
conditions. \citet{MontvilleMatthews2001}, who studied the effect of
temperature with different pH values, concluded that the growth rate
increases with temperature, reaching a maximum at 40$^\circ$C to decay
afterward. The behavior of the growth rate is similar for different
values of pH. Similar conclusions can be found in the literature,
however, in our work the results are fairly different. The effect of
temperature on the growth rate is not as it was shown in previous
studies. Secondary models show a very continuous line of increasing
growth rate values with temperature or pH changes [\citet{MontvilleMatthews2001}; \citet{mckellar20042}]. In contrast, in our
work, for a fixed pH value the effect of temperature is not so smooth,
and the growth rate shows oscillations that have not been described in
the literature with any secondary model as far as we know. Moreover,
the maximum growth rates are achieved at different levels of
temperature when the pH values vary, differing from the CAPM models.

Specifically, when pH is about 7.4, the temperature must be diminished
to decrease the growth rate of \textit{Listeria}, but if pH is about
4.5, then the temperature must be increased to decrease growth rates.
Therefore, consequences in terms of food conservation vary regarding
their respective pH. For example, in fruits and vegetables, which
present in general low pH values, it is convenient to increase
temperature; on the contrary, in dairy products, biscuits, chocolate
and eggs it is convenient to decrease temperatures.

It is generally accepted that predictions of the response within that
range can be made by interpolation. Usually, a three-dimensional space
could be constructed with the ranges of the three parameters studied.
The interpolation region of the combinations tested is called the
minimum convex polyhedron [\citet{baranyi1996effects}] and it is used
to make predictions. However, \citet{baranyi1996effects} noted that
this is not always a good approach. These authors reported a prediction
of an optimal growth for \textit{Salmonella} under approximately 2\%
NaCl, which is not correct for that microorganism. Additionally, \citet{davey1989predictive} noted that polynomial equations did not have a
consistent form across a range of bacterial growth data and that such
models appeared to lack of universality. That is, the coefficient
values of polynomial models are very data dependent. In this work, we
have implemented two kinds of predictions which were not widely used in
the literature but which are of greater relevance in the context of the
microbiology. We have shown that predictions out of sample are very
accurate, being a good alternative to the use of polynomials of
different orders (2nd or 3rd order) and response surfaces for
predictive microbiology.
%Baranyi et al. (1996) concluded that it is inappropriate to introduce
%parameters in a model or to increase
%the order of the polynomial simply to improve the goodness-of-fit of
%the model.

A restriction in the models, as assumed here, is that we suppose that
data are equally spaced in time. Although this is typically the case in
Petri dish or in optical density experiments, this may not be true with
more general populations. In the case of irregularly spaced data,
differential equation models with diffusion type approximations with
the neural network models for the growth functions may be considered
[see \citet{donnet2010bayesian}].

Finally, alternative approximations to the hierarchical neural network
models for growth functions may be considered as spline methods from a
classical point of view, functional data analysis or Gaussian process
approximations.

\begin{supplement}
\sname{Supplement A}
\stitle{Code for the NN and GNN models\\}
\slink[doi]{10.1214/14-AOAS720SUPPA}
\sdatatype{.zip}
\sfilename{aoas720\_suppA.zip}
\sdescription{The file contains two programs,
\texttt{NN model.odc} for running the neural network model and \texttt
{GNN model.odc} for running the Gompertz neural network model.}
\end{supplement}

\begin{supplement}
\sname{Supplement B}
\stitle{Data sets}
\slink[doi]{10.1214/14-AOAS720SUPPB}
\sdatatype{.xls}
\sfilename{aoas720\_suppB.xls}
\sdescription{The file \texttt{data.xls} contains 10 replications in
6 groups of bacteria under the environmental conditions outlined in
Figure~\ref{figfactors}.}
\end{supplement}


%suskaldyti doi

% imsref loaded by linak, 2014-07-11 10:21:10
% imsref loaded by linak, 2014-07-11 11:19:02
% imsref loaded by linak, 2014-07-11 11:45:34
%
% imsref loaded by linak, 2014-07-15 09:21:48
% imsref loaded by linak, 2014-07-15 09:30:53
% imsref loaded by linak, 2014-07-15 09:32:30

\printaddresses


\begin{thebibliography}{58}

%b1 #&#
\bibitem[\protect\citeauthoryear{Andrieu, de~Freitas and
Doucet}{2001}]{andrieu2001robust}
%
\begin{barticle}[pbm]
\bauthor{\bsnm{Andrieu},~\bfnm{C.}\binits{C.}},
\bauthor{\bparticle{de} \bsnm{Freitas},~\bfnm{N.}\binits{N.}} \AND
\bauthor{\bsnm{Doucet},~\bfnm{A.}\binits{A.}}
(\byear{2001}).
\btitle{Robust full Bayesian learning for radial basis networks}.
\bjournal{Neural Comput.}
\bvolume{13}
\bpages{2359--2407}.
\bid{doi={10.1162/089976601750541831}, issn={0899-7667}, pmid={11571002}}
\end{barticle}
%
\bptok{imsref}%
% NOT OUTPUTED:
% issn = 0899-7667
% number = 10
% fjournal = Neural computation
\endbibitem

%b2 #&#
\bibitem[\protect\citeauthoryear{Augustin and
Carlier}{2000}]{augustin2000modelling}
%
\begin{barticle}[pbm]
\bauthor{\bsnm{Augustin},~\bfnm{J.~C.}\binits{J.~C.}} \AND
\bauthor{\bsnm{Carlier},~\bfnm{V.}\binits{V.}}
(\byear{2000}).
\btitle{Modelling the growth rate of \textit{Listeria monocytogenes} with a
multiplicative type model including interactions between environmental factors}.
\bjournal{Int. J. Food Microbiol.}
\bvolume{56}
\bpages{53--70}.
\bid{issn={0168-1605}, pii={S0168-1605(00)00224-5}, pmid={10857925}}
\end{barticle}
%
\bptok{imsref}%
% NOT OUTPUTED:
% issn = 0168-1605
% number = 1
% fjournal = International journal of food microbiology
\endbibitem

%b3 #&#
\bibitem[\protect\citeauthoryear{Baranyi et~al.}{1996}]{baranyi1996effects}
%
\begin{barticle}[author]
\bauthor{\bsnm{Baranyi},~\bfnm{J.}\binits{J.}},
\bauthor{\bsnm{Ross},~\bfnm{T.}\binits{T.}},
\bauthor{\bsnm{McMeekin},~\bfnm{T.~A.}\binits{T.~A.}} \AND
\bauthor{\bsnm{Roberts},~\bfnm{T.~A.}\binits{T.~A.}}
(\byear{1996}).
\btitle{Effects of parameterization on the performance of empirical
models used in predictive microbiology}.
\bjournal{Food Microbiology}
\bvolume{13}
\bpages{83--91}.
\end{barticle}
%
\bptok{imsref}%
\endbibitem

%b4 #&#
\bibitem[\protect\citeauthoryear{Brooks and Gelman}{1998}]{brooks1998general}
%
\begin{barticle}[mr]
\bauthor{\bsnm{Brooks},~\bfnm{Stephen~P.}\binits{S.~P.}} \AND
\bauthor{\bsnm{Gelman},~\bfnm{Andrew}\binits{A.}}
(\byear{1998}).
\btitle{General methods for monitoring convergence of iterative simulations}.
\bjournal{J. Comput. Graph. Statist.}
\bvolume{7}
\bpages{434--455}.
\bid{doi={10.2307/1390675}, issn={1061-8600}, mr={1665662}}
\end{barticle}
%
\bptok{imsref}%
% NOT OUTPUTED:
% issn = 1061-8600
% url = http://dx.doi.org/10.2307/1390675
% number = 4
% fjournal = Journal of Computational and Graphical Statistics
\endbibitem

%b5 #&#
\bibitem[\protect\citeauthoryear{CAC}{1996}]{CAC1996}
%
\begin{bmisc}[author]
\bauthor{\bsnm{{CAC}}}
(\byear{1996}).
\bhowpublished{Principles and guidelines for the application of
microbiological risk assessment.
Technical report, Codex Alimentarius Commission, CX/FH 96/10 FAO, Rome, Italy.}
\end{bmisc}
%
\bptok{imsref}%
% NOT OUTPUTED:
\endbibitem

%b6 #&#
\bibitem[\protect\citeauthoryear{CEC}{2002}]{CEC2002}
%
\begin{bmisc}[author]
\bauthor{\bsnm{CEC}}
(\byear{2002}).
\bhowpublished{Regulation (EC) 178/2002 of 28 January 2002 on laying down
the general principles and requirements of food law, establishing the
European Food Safety Authority and laying down procedures in matters of
food safety. OJ L 31.
Technical report, Commission of the European Communities.}
\end{bmisc}
%
\bptok{imsref}%
% NOT OUTPUTED:
\endbibitem

%b7 #&#
\bibitem[\protect\citeauthoryear{CEC}{2005}]{CEC2005}
%
\begin{bmisc}[author]
\bauthor{\bsnm{CEC}}
(\byear{2005}).
\btitle{Regulation (EC) 2073/2005 of 15 November 2005 on
microbiological criteria for foodstuffs. OJ L 338.
Technical report, Commission of the European Communities.}
\end{bmisc}
%
\bptok{imsref}%
% NOT OUTPUTED:
\endbibitem

%b8 #&#
\bibitem[\protect\citeauthoryear{Celeux et~al.}{2006}]{celeux2006deviance}
%
\begin{barticle}[mr]
\bauthor{\bsnm{Celeux},~\bfnm{G.}\binits{G.}},
\bauthor{\bsnm{Forbes},~\bfnm{F.}\binits{F.}},
\bauthor{\bsnm{Robert},~\bfnm{C.~P.}\binits{C.~P.}} \AND
\bauthor{\bsnm{Titterington},~\bfnm{D.~M.}\binits{D.~M.}}
(\byear{2006}).
\btitle{Deviance information criteria for missing data models}.
\bjournal{Bayesian Anal.}
\bvolume{1}
\bpages{651--673 (electronic)}.
\bid{doi={10.1214/06-BA122}, issn={1931-6690}, mr={2282197}}
\end{barticle}
%
\bptok{imsref}%
% NOT OUTPUTED:
% issn = 1931-6690
% url = http://dx.doi.org/10.1214/06-BA122
% number = 4
% fjournal = Bayesian Analysis
\endbibitem

%b9 #&#
\bibitem[\protect\citeauthoryear{Davey}{1989}]{davey1989predictive}
%
\begin{barticle}[author]
\bauthor{\bsnm{Davey},~\bfnm{K.~R.}\binits{K.~R.}}
(\byear{1989}).
\btitle{A predictive model for combined temperature and water activity
on microbial growth during the growth phase}.
\bjournal{J. Appl. Microbiol.}
\bvolume{67}
\bpages{483--488}.
\end{barticle}
%
\bptok{imsref}%
\endbibitem

%b10 #&#
\bibitem[\protect\citeauthoryear{Davidson}{2001}]{davidson2001}
%
\begin{bincollection}[author]
\bauthor{\bsnm{Davidson},~\bfnm{P.~M.}\binits{P.~M.}}
(\byear{2001}).
\btitle{Chemical preservatives and natural antimicrobial compounds}.
In \bbooktitle{Food Microbiology: Fundamentals and Frontiers}.
\bpublisher{ASM Press},
\blocation{Washington, DC}.
\end{bincollection}
%
\bptok{imsref}%
\endbibitem

%b11 #&#
\bibitem[\protect\citeauthoryear{Donnet, Foulley and
Samson}{2010}]{donnet2010bayesian}
%
\begin{barticle}[mr]
\bauthor{\bsnm{Donnet},~\bfnm{Sophie}\binits{S.}},
\bauthor{\bsnm{Foulley},~\bfnm{Jean-Louis}\binits{J.-L.}} \AND
\bauthor{\bsnm{Samson},~\bfnm{Adeline}\binits{A.}}
(\byear{2010}).
\btitle{Bayesian analysis of growth curves using mixed models defined
by stochastic differential equations}.
\bjournal{Biometrics}
\bvolume{66}
\bpages{733--741}.
\bid{doi={10.1111/j.1541-0420.2009.01342.x}, issn={0006-341X}, mr={2758209}}
\end{barticle}
%
\bptok{imsref}%
% NOT OUTPUTED:
% issn = 0006-341X
% url = http://dx.doi.org/10.1111/j.1541-0420.2009.01342.x
% number = 3
% fjournal = Biometrics. Journal of the International Biometric Society
\endbibitem

%b12 #&#
\bibitem[\protect\citeauthoryear{Eduardo et~al.}{2011}]{eduardo2011analyzing}
%
\begin{barticle}[author]
\bauthor{\bsnm{Eduardo},~\bfnm{A.~J.~S.}\binits{A.~J.~S.}},
\bauthor{\bsnm{Quinto},~\bfnm{E.~J.}\binits{E.~J.}},
\bauthor{\bsnm{Castro},~\bfnm{M.~J.}\binits{M.~J.}} \AND
\bauthor{\bsnm{Mora},~\bfnm{M.~T.}\binits{M.~T.}}
(\byear{2011}).
\btitle{Analyzing the time-to-detection--generation time relationship
of Escherichia coli}.
\bjournal{CyTA-Journal of Food}
\bvolume{9}
\bpages{271--277}.
\end{barticle}
%
\bptok{imsref}%
\endbibitem

%b13 #&#
\bibitem[\protect\citeauthoryear{{FAO/WHO}}{1995}]{FAO1995}
%
\begin{bincollection}[author]
\bauthor{\bsnm{{FAO/WHO}}}
(\byear{1995}).
\btitle{Uncertainty and variability in the risk assessment process}.
In \bbooktitle{Application of Risk Analysis to Food Standards Issues}.
\bpublisher{Agriculture and Consumer Protection Dept.},
\blocation{Geneva}.
\bnote{Available at \url{http://www.fao.org/docrep/008/ae922e/ae922e00.htm}}.
\end{bincollection}
%
\bptok{imsref}%
\endbibitem

%b14 #&#
\bibitem[\protect\citeauthoryear{Fennema and Tannenbaum}{1996}]{fennema1996introduction}
%
\begin{bincollection}[author]
\bauthor{\bsnm{Fennema},~\bfnm{Owen~R.}\binits{O.~R.}} \AND
\bauthor{\bsnm{Tannenbaum},~\bfnm{Steven~R.}\binits{S.~R.}}
(\byear{1996}).
\btitle{Introduction to food chemistry}.
In \bbooktitle{Food Chemistry},
\bedition{3th} ed.
(\beditor{\bfnm{O.~R.}\binits{O.~R.}~\bsnm{Fennema}}, ed.)
\bpages{1--16}.
\bpublisher{Dekker}, \blocation{New York}.
\end{bincollection}
%
\bptok{imsref}%
\endbibitem

%b15 #&#
\bibitem[\protect\citeauthoryear{Fine}{1999}]{fine1999feedforward}
%
\begin{bbook}[mr]
\bauthor{\bsnm{Fine},~\bfnm{Terrence~L.}\binits{T.~L.}}
(\byear{1999}).
\btitle{Feedforward Neural Network Methodology}.
\bpublisher{Springer},
\blocation{New York}.
\bid{mr={1691898}}
\end{bbook}
%
\bptok{imsref}%
% NOT OUTPUTED:
% isbn = 0-387-98745-2
% fpage = xvi+340
\endbibitem

%b16 #&#
\bibitem[\protect\citeauthoryear{Garc\'ia-Gimeno et~al.}{2002}]{garcia2002improving}
%
\begin{barticle}[author]
\bauthor{\bsnm{Garc\'ia-Gimeno},~\bfnm{R.~M.}\binits{R.~M.}},
\bauthor{\bsnm{Herv{\'a}s-Mart\'inez},~\bfnm{C.}\binits{C.}} \betal{et~al.}
(\byear{2002}).
\btitle{{Improving artificial neural networks with a pruning
methodology and genetic algorithms for their application in microbial
growth prediction in food.}}
\bjournal{Int. J. Food Microbiol.}
\bvolume{72}
\bpages{19--30}.
\end{barticle}
%
\bptok{imsref}%
\endbibitem

%b17 #&#
\bibitem[\protect\citeauthoryear{Geeraerd et~al.}{1998}]{geeraerd1998application}
%
\begin{barticle}[pbm]
\bauthor{\bsnm{Geeraerd},~\bfnm{A.~H.}\binits{A.~H.}},
\bauthor{\bsnm{Herremans},~\bfnm{C.~H.}\binits{C.~H.}},
\bauthor{\bsnm{Cenens},~\bfnm{C.}\binits{C.}} \AND
\bauthor{\bsnm{Impe},~\bfnm{J.~F.~Van}\binits{J.~F.~V.}}
(\byear{1998}).
\btitle{Application of artificial neural networks as a non-linear
modular modeling technique to describe bacterial growth in chilled food
products}.
\bjournal{Int. J. Food Microbiol.}
\bvolume{44}
\bpages{49--68}.
\bid{issn={0168-1605}, pii={S0168-1605(98)00127-5}, pmid={9849784}}
\end{barticle}
%
\bptok{imsref}%
% NOT OUTPUTED:
% issn = 0168-1605
% number = 1-2
% fjournal = International journal of food microbiology
\endbibitem

%b18 #&#
\bibitem[\protect\citeauthoryear{Gelfand}{1996}]{gelfand1996model}
%
\begin{bincollection}[mr]
\bauthor{\bsnm{Gelfand},~\bfnm{Alan~E.}\binits{A.~E.}}
(\byear{1996}).
\btitle{Model determination using sampling-based methods}.
In \bbooktitle{Markov Chain {M}onte {C}arlo in Practice}
(\beditor{\bfnm{W.}\binits{W.}~\bsnm{Gilks}},
\beditor{\bfnm{S.}\binits{S.}~\bsnm{Richardson}}
\AND
\beditor{\bfnm{D.}\binits{D.}~\bsnm{Spiegelhalter}}, eds.)
\bpages{145--161}.
\bpublisher{Chapman \& Hall},
\blocation{London}.
\bid{mr={1397969}}
\end{bincollection}
%
\bptok{imsref}%
\endbibitem

%b19 #&#
\bibitem[\protect\citeauthoryear{Gelfand and Ghosh}{1998}]{gelfand1998model}
%
\begin{barticle}[mr]
\bauthor{\bsnm{Gelfand},~\bfnm{Alan~E.}\binits{A.~E.}} \AND
\bauthor{\bsnm{Ghosh},~\bfnm{Sujit~K.}\binits{S.~K.}}
(\byear{1998}).
\btitle{Model choice: A minimum posterior predictive loss approach}.
\bjournal{Biometrika}
\bvolume{85}
\bpages{1--11}.
\bid{doi={10.1093/biomet/85.1.1}, issn={0006-3444}, mr={1627258}}
\end{barticle}
%
\bptok{imsref}%
% NOT OUTPUTED:
% issn = 0006-3444
% url = http://dx.doi.org/10.1093/biomet/85.1.1
% number = 1
% coden = BIOKAX
% fjournal = Biometrika
\endbibitem

%b20 #&#
\bibitem[\protect\citeauthoryear{Gilks, Richardson and
Spiegelhalter}{1996}]{gilks1996markov}
%
\begin{bbook}[mr]
\bauthor{\bsnm{Gilks},~\bfnm{W.~R.}\binits{W.~R.}},
 \bauthor{\bsnm{Richardson},~\bfnm{S.}\binits{S.}} \AND
 \bauthor{\bsnm{Spiegelhalter},~\bfnm{D.~J.}\binits{D.~J.}}
(\byear{1996}).
\btitle{Markov Chain {M}onte {C}arlo in Practice}.
\bpublisher{Chapman \& Hall},
\blocation{London}.
\bid{doi={10.1007/978-1-4899-4485-6}, mr={1397966}}
\end{bbook}
%
\bptok{imsref}%
% NOT OUTPUTED:
% isbn = 0-412-05551-1
% url = http://dx.doi.org/10.1007/978-1-4899-4485-6
% fpage = xviii+486
\endbibitem

%b21 #&#
\bibitem[\protect\citeauthoryear{Gompertz}{1825}]{gompertz1825on}
%
\begin{barticle}[author]
\bauthor{\bsnm{Gompertz},~\bfnm{Benjamin}\binits{B.}}
(\byear{1825}).
\btitle{{On the nature of the function expressive of the law of human
mortality, and on a new mode of determining the value of life contingencies}}.
\bjournal{Philos. Trans. R. Soc. Lond.}
\bvolume{115}
\bpages{513--583}.
\end{barticle}
%
\bptok{imsref}%
\endbibitem

%b22 #&#
\bibitem[\protect\citeauthoryear{Hajmeer, Basheer and
Najjar}{1997}]{hajmeer1997computational}
%
\begin{barticle}[pbm]
\bauthor{\bsnm{Hajmeer},~\bfnm{M.~N.}\binits{M.~N.}},
\bauthor{\bsnm{Basheer},~\bfnm{I.~A.}\binits{I.~A.}} \AND
\bauthor{\bsnm{Najjar},~\bfnm{Y.~M.}\binits{Y.~M.}}
(\byear{1997}).
\btitle{Computational neural networks for predictive microbiology. II.
Application to microbial growth}.
\bjournal{Int. J. Food Microbiol.}
\bvolume{34}
\bpages{51--66}.
\bid{issn={0168-1605}, pii={S0168160596011695}, pmid={9029255}}
\end{barticle}
%
\bptok{imsref}%
% NOT OUTPUTED:
% issn = 0168-1605
% number = 1
% fjournal = International journal of food microbiology
\endbibitem

%b23 #&#
\bibitem[\protect\citeauthoryear{ICMSF}{1980}]{ICMSF1980}
%
\begin{bbook}[author]
\bauthor{\bsnm{ICMSF}}
(\byear{1980}).
\btitle{Microbial Ecology of Foods}.
\bseries{Factors Affecting the Life and Death of Microorganisms}
\bvolume{1}.
\bpublisher{Academic Press},
\blocation{London}.
\end{bbook}
%
\bptok{imsref}%
\endbibitem

%b24 #&#
\bibitem[\protect\citeauthoryear{{IOM, Institute of Medicine of
National Academies}}{2010}]{IOM2010}
%
\begin{bincollection}[author]
\bauthor{\bsnm{{IOM, Institute of Medicine of National Academies}}}
(\byear{2010}).
\btitle{Preservation and physical property roles of sodium in foods}.
In \bbooktitle{Strategies to Reduce Sodium Intake in the United States.
Committee on Strategies to Reduce Sodium Intake}.
\bpublisher{National Academies Press},
\blocation{Washington, DC}.
\bnote{Available at \url{http://www.ncbi.nlm.nih.gov/books/NBK50952/}}.
\end{bincollection}
%
\bptok{imsref}%
\endbibitem

%b25 #&#
\bibitem[\protect\citeauthoryear{Lavine and West}{1992}]{lavine1992bayesian}
%
\begin{barticle}[mr]
\bauthor{\bsnm{Lavine},~\bfnm{Michael}\binits{M.}} \AND
\bauthor{\bsnm{West},~\bfnm{Mike}\binits{M.}}
(\byear{1992}).
\btitle{A {B}ayesian method for classification and discrimination}.
\bjournal{Canad. J. Statist.}
\bvolume{20}
\bpages{451--461}.
\bid{doi={10.2307/3315614}, issn={0319-5724}, mr={1208356}}
\end{barticle}
%
\bptok{imsref}%
% NOT OUTPUTED:
% issn = 0319-5724
% url = http://dx.doi.org/10.2307/3315614
% number = 4
% fjournal = The Canadian Journal of Statistics. La Revue Canadienne de
%Statistique
\endbibitem

%b26 #&#
\bibitem[\protect\citeauthoryear{Lee}{2004}]{lee2004bayesian}
%
\begin{bbook}[author]
\bauthor{\bsnm{Lee},~\bfnm{H.~K.~H.}\binits{H.~K.~H.}}
(\byear{2004}).
\btitle{Bayesian Nonparametrics via Neural Networks}.
\bpublisher{SIAM},
\blocation{Philadelphia, PA}.
\end{bbook}
%
\bptok{imsref}%
\endbibitem

%b27 #&#
\bibitem[\protect\citeauthoryear{Leistner}{2000}]{leistner2000basic}
%
\begin{barticle}[author]
\bauthor{\bsnm{Leistner},~\bfnm{Lothar}\binits{L.}}
(\byear{2000}).
\btitle{Basic aspects of food preservation by hurdle technology}.
\bjournal{Int. J. Food Microbiol.}
\bvolume{55}
\bpages{181--186}.
\end{barticle}
%
\bptok{imsref}%
\endbibitem

%b28 #&#
\bibitem[\protect\citeauthoryear{MacKay}{1995}]{mackay1995bayesian}
%
\begin{bmisc}[author]
\bauthor{\bsnm{MacKay},~\bfnm{D.~J.~C.}\binits{D.~J.~C.}}
(\byear{1995}).
\bhowpublished{Bayesian methods for neural networks: Theory and applications.
Technical report, Cavendish Laboratory, Cambridge Univ., Cambridge.}
\end{bmisc}
%
\bptok{imsref}%
% NOT OUTPUTED:
\endbibitem

%b29 #&#
\bibitem[\protect\citeauthoryear{McClure et~al.}{1993}]{mcclure1993predictive}
%
\begin{barticle}[author]
\bauthor{\bsnm{McClure},~\bfnm{P.~J.}\binits{P.~J.}},
\bauthor{\bsnm{Baranyi},~\bfnm{J.}\binits{J.}},
\bauthor{\bsnm{Boogard},~\bfnm{E.}\binits{E.}},
\bauthor{\bsnm{Kelly},~\bfnm{T.~M.}\binits{T.~M.}} \AND
\bauthor{\bsnm{Roberts},~\bfnm{T.~A.}\binits{T.~A.}}
(\byear{1993}).
\btitle{A~predictive model for the combined effect of pH, sodium
chloride and storage temperature on the growth of \textit{Brochothrix thermosphacta}}.
\bjournal{Int. J. Food Microbiol.}
\bvolume{19}
\bpages{161--178}.
\end{barticle}
%
\bptok{imsref}%
\endbibitem

%b30 #&#
\bibitem[\protect\citeauthoryear{McKellar and Lu}{2004}]{mckellar20042}
%
\begin{bbook}[author]
\bauthor{\bsnm{McKellar},~\bfnm{R.~C.}\binits{R.~C.}} \AND
\bauthor{\bsnm{Lu},~\bfnm{X.}\binits{X.}}
(\byear{2004}).
\btitle{Modeling Microbial Responses in Food}.
\bpublisher{CRC Press},
\blocation{Boca Raton, FL}.
\end{bbook}
%
\bptok{imsref}%
\endbibitem

%b31 #&#
\bibitem[\protect\citeauthoryear{McMeekin et~al.}{1987}]{mcmeekin1987model}
%
\begin{barticle}[author]
\bauthor{\bsnm{McMeekin},~\bfnm{T.~A.}\binits{T.~A.}},
\bauthor{\bsnm{Chandler},~\bfnm{R.~E.}\binits{R.~E.}},
\bauthor{\bsnm{Doe},~\bfnm{P.~E.}\binits{P.~E.}},
\bauthor{\bsnm{Garland},~\bfnm{C.~D.}\binits{C.~D.}},
\bauthor{\bsnm{Olley},~\bfnm{J.}\binits{J.}},
\bauthor{\bsnm{Putro},~\bfnm{S.}\binits{S.}} \AND
\bauthor{\bsnm{Ratkowsky},~\bfnm{D.~A.}\binits{D.~A.}}
(\byear{1987}).
\btitle{Model for combined effect of temperature and salt
concentration/water activity on the growth rate of \textit{Staphylococcus xylosus}}.
\bjournal{J. Appl. Microbiol.}
\bvolume{62}
\bpages{543--550}.
\end{barticle}
%
\bptok{imsref}%
\endbibitem

%b32 #&#
\bibitem[\protect\citeauthoryear{Miles et~al.}{1997}]{miles1997development}
%
\begin{barticle}[pbm]
\bauthor{\bsnm{Miles},~\bfnm{D.~W.}\binits{D.~W.}},
\bauthor{\bsnm{Ross},~\bfnm{T.}\binits{T.}},
\bauthor{\bsnm{Olley},~\bfnm{J.}\binits{J.}} \AND
\bauthor{\bsnm{McMeekin},~\bfnm{T.~A.}\binits{T.~A.}}
(\byear{1997}).
\btitle{Development and evaluation of a predictive model for the effect
of temperature and water activity on the growth rate of \textit{Vibrio
parahaemolyticus}}.
\bjournal{Int. J. Food Microbiol.}
\bvolume{38}
\bpages{133--142}.
\bid{issn={0168-1605}, pii={S0168-1605(97)00100-1}, pmid={9506279}}
\end{barticle}
%
\bptok{imsref}%
% NOT OUTPUTED:
% issn = 0168-1605
% number = 2-3
% fjournal = International journal of food microbiology
\endbibitem

%b33 #&#
\bibitem[\protect\citeauthoryear{Montville and
Matthews}{2001}]{MontvilleMatthews2001}
%
\begin{bincollection}[author]
\bauthor{\bsnm{Montville},~\bfnm{T.~J.}\binits{T.~J.}} \AND
\bauthor{\bsnm{Matthews},~\bfnm{K.~R.}\binits{K.~R.}}
(\byear{2001}).
\btitle{Principles which influence microbial growth, survival, and
death of foods}.
In \bbooktitle{Food Microbiology: Fundamentals and Frontiers}.
\bpublisher{ASM Press},
\blocation{Washington, DC}.
\end{bincollection}
%
\bptok{imsref}%
\endbibitem

%b34 #&#
\bibitem[\protect\citeauthoryear{Montville and Matthews}{2005}]{montville2005food}
%
\begin{bbook}[author]
\bauthor{\bsnm{Montville},~\bfnm{Thomas~J.}\binits{T.~J.}} and
\bauthor{\bsnm{Matthews},~\bfnm{Karl~R.}\binits{K.~R.}}
(\byear{2005}).
\btitle{Food Microbiology: An Introduction},
\bedition{1st} ed.
\bpublisher{ASM Press},
\blocation{Washington, DC}.
\end{bbook}
%
\bptok{imsref}%
\endbibitem

%b35 #&#
\bibitem[\protect\citeauthoryear{M{\"{u}}ller and Insua}{1998}]{muller1998iba}
%
\begin{barticle}[author]
\bauthor{\bsnm{M{\"{u}}ller},~\bfnm{Peter}\binits{P.}} \AND
\bauthor{\bsnm{Insua},~\bfnm{David~Rios.}\binits{D.~R.}}
(\byear{1998}).
\btitle{Issues in Bayesian analysis of neural network models}.
\bjournal{Neural Comput.}
\bvolume{10}
\bpages{749--770}.
\end{barticle}
%
\bptok{imsref}%
\endbibitem

%b36 #&#
\bibitem[\protect\citeauthoryear{{NACMCF}}{1997}]{NACMCF1997}
%
\begin{bmisc}[author]
\bauthor{\bsnm{{NACMCF}}}
(\byear{1997}).
\bhowpublished{Principles and guidelines for the application of
microbiological risk assessment.
Technical report, USA Dept. Agriculture, Food Safety and
Inspection Service, Washington, DC.}
\end{bmisc}
%
\bptok{imsref}%
% NOT OUTPUTED:
\endbibitem

%b37 #&#
\bibitem[\protect\citeauthoryear{Neal}{1996}]{neal1996bayesian}
%
\begin{bbook}[author]
\bauthor{\bsnm{Neal},~\bfnm{R.~M.}\binits{R.~M.}}
(\byear{1996}).
\btitle{Bayesian Learning for Neural Networks}.
\bpublisher{Springer},
\blocation{Berlin}.
\end{bbook}
%
\bptok{imsref}%
\endbibitem

%b38 #&#
\bibitem[\protect\citeauthoryear{Palacios et~al.}{2014a}]{supplementA}
\begin{bmisc}[auto]
\bauthor{\bsnm{Palacios},~\bfnm{A. Paula}\binits{A.~P.}},
\bauthor{\bsnm{Mar\'{\i}n},~\bfnm{J. Miguel}\binits{J.~M.}},
\bauthor{\bsnm{Quinto},~\bfnm{Emiliano J.}\binits{E.~J.}} \AND
\bauthor{\bsnm{Wiper},~\bfnm{Michael P.}\binits{M.~P.}}
(\byear{2014a}).
\bhowpublished{Supplement to ``Bayesian modeling of bacterial growth
for multiple populations.''
DOI:\doiurl{10.1214/14-AOAS720SUPPA}}.
\end{bmisc}
\bptok{imsref}%
\endbibitem

%b39 #&#
\bibitem[\protect\citeauthoryear{Palacios et~al.}{2014b}]{supplementB}
\begin{bmisc}[auto]
\bauthor{\bsnm{Palacios},~\bfnm{A. Paula}\binits{A.~P.}},
\bauthor{\bsnm{Mar\'{\i}n},~\bfnm{J. Miguel}\binits{J.~M.}},
\bauthor{\bsnm{Quinto},~\bfnm{Emiliano J.}\binits{E.~J.}} \AND
\bauthor{\bsnm{Wiper},~\bfnm{Michael P.}\binits{M.~P.}}
(\byear{2014b}).
\bhowpublished{Supplement to ``Bayesian modeling of bacterial growth
for multiple populations.''
DOI:\doiurl{10.1214/14-AOAS720SUPPB}}.
\end{bmisc}
\bptok{imsref}%
\endbibitem

%b40 #&#
\bibitem[\protect\citeauthoryear{Potter and Hotchkiss}{1998}]{potter1998food}
%
\begin{bbook}[author]
\bauthor{\bsnm{Potter},~\bfnm{Norman~N.}\binits{N.~N.}} \AND
\bauthor{\bsnm{Hotchkiss},~\bfnm{Joseph~H.}\binits{J.~H.}}
(\byear{1998}).
\btitle{Food Science}.
\bpublisher{Springer},
\blocation{Berlin}.
\end{bbook}
%
\bptok{imsref}%
\endbibitem

%b41 #&#
\bibitem[\protect\citeauthoryear{Pouillot et~al.}{2003}]{pouillot2003estimation}
%
\begin{barticle}[pbm]
\bauthor{\bsnm{Pouillot},~\bfnm{R{\'{e}}gis}\binits{R.}},
\bauthor{\bsnm{Albert},~\bfnm{Isabelle}\binits{I.}},
\bauthor{\bsnm{Cornu},~\bfnm{Marie}\binits{M.}} \AND
\bauthor{\bsnm{Denis},~\bfnm{Jean~Baptiste}\binits{J.~B.}}
(\byear{2003}).
\btitle{Estimation of uncertainty and variability in bacterial growth
using Bayesian inference. Application to \textit{Listeria monocytogenes}}.
\bjournal{Int. J. Food Microbiol.}
\bvolume{81}
\bpages{87--104}.
\bid{issn={0168-1605}, pii={S0168160502001927}, pmid={12457583}}
\end{barticle}
%
\bptok{imsref}%
% NOT OUTPUTED:
% issn = 0168-1605
% number = 2
% fjournal = International journal of food microbiology
\endbibitem

%b42 #&#
\bibitem[\protect\citeauthoryear{Ratkowsky
et~al.}{1982}]{ratkowsky1982relationship}
%
\begin{barticle}[pbm]
\bauthor{\bsnm{Ratkowsky},~\bfnm{D.~A.}\binits{D.~A.}},
\bauthor{\bsnm{Olley},~\bfnm{J.}\binits{J.}},
\bauthor{\bsnm{McMeekin},~\bfnm{T.~A.}\binits{T.~A.}} \AND
\bauthor{\bsnm{Ball},~\bfnm{A.}\binits{A.}}
(\byear{1982}).
\btitle{Relationship between temperature and growth rate of bacterial cultures}.
\bjournal{J. Bacteriol.}
\bvolume{149}
\bpages{1--5}.
\bid{issn={0021-9193}, pmcid={216584}, pmid={7054139}}
\end{barticle}
%
\bptok{imsref}%
% NOT OUTPUTED:
% issn = 0021-9193
% number = 1
% fjournal = Journal of bacteriology
\endbibitem

%b43 #&#
\bibitem[\protect\citeauthoryear{Robert and
Mengersen}{1999}]{robert1999reparameterisation}
%
\begin{barticle}[author]
\bauthor{\bsnm{Robert},~\bfnm{C.~P.}\binits{C.~P.}} \AND
\bauthor{\bsnm{Mengersen},~\bfnm{K.~L.}\binits{K.~L.}}
(\byear{1999}).
\btitle{Reparameterisation issues in mixture modelling and their
bearing on MCMC algorithms}.
\bjournal{Comput. Statist. Data Anal.}
\bvolume{29}
\bpages{325--344}.
\end{barticle}
%
\bptok{imsref}%
\endbibitem

%b44 #&#
\bibitem[\protect\citeauthoryear{Roeder and
Wasserman}{1997}]{roeder1997practical}
%
\begin{barticle}[mr]
\bauthor{\bsnm{Roeder},~\bfnm{Kathryn}\binits{K.}} \AND
\bauthor{\bsnm{Wasserman},~\bfnm{Larry}\binits{L.}}
(\byear{1997}).
\btitle{Practical {B}ayesian density estimation using mixtures of normals}.
\bjournal{J. Amer. Statist. Assoc.}
\bvolume{92}
\bpages{894--902}.
\bid{doi={10.2307/2965553}, issn={0162-1459}, mr={1482121}}
\end{barticle}
%
\bptok{imsref}%
% NOT OUTPUTED:
% issn = 0162-1459
% url = http://dx.doi.org/10.2307/2965553
% number = 439
% coden = JSTNAL
% fjournal = Journal of the American Statistical Association
\endbibitem

%b45 #&#
\bibitem[\protect\citeauthoryear{Ross and Dalgaard}{2004}]{rossdalgaard2004}
%
\begin{bincollection}[author]
\bauthor{\bsnm{Ross},~\bfnm{T.}\binits{T.}} \AND
\bauthor{\bsnm{Dalgaard},~\bfnm{P.}\binits{P.}}
(\byear{2004}).
\btitle{Secondary models}.
In \bbooktitle{Modeling Microbial Responses in Food}.
\bpublisher{CRC Press},
\blocation{Boca Raton, FL}.
\end{bincollection}
%
\bptok{imsref}%
\endbibitem

%b46 #&#
\bibitem[\protect\citeauthoryear{Ross and McMeekin}{1994}]{ross1994predictive}
%
\begin{barticle}[pbm]
\bauthor{\bsnm{Ross},~\bfnm{T.}\binits{T.}} \AND
\bauthor{\bsnm{McMeekin},~\bfnm{T.~A.}\binits{T.~A.}}
(\byear{1994}).
\btitle{Predictive microbiology}.
\bjournal{Int. J. Food Microbiol.}
\bvolume{23}
\bpages{241--264}.
\bid{issn={0168-1605}, pmid={7873329}}
\end{barticle}
%
\bptok{imsref}%
% NOT OUTPUTED:
% issn = 0168-1605
% number = 3-4
% fjournal = International journal of food microbiology
\endbibitem

%b47 #&#
\bibitem[\protect\citeauthoryear{Rosso et~al.}{1995}]{rosso1995convenient}
%
\begin{barticle}[pbm]
\bauthor{\bsnm{Rosso},~\bfnm{L.}\binits{L.}},
\bauthor{\bsnm{Lobry},~\bfnm{J.~R.}\binits{J.~R.}},
\bauthor{\bsnm{Bajard},~\bfnm{S.}\binits{S.}} \AND
\bauthor{\bsnm{Flandrois},~\bfnm{J.~P.}\binits{J.~P.}}
(\byear{1995}).
\btitle{Convenient model to describe the combined effects of
temperature and pH on microbial growth}.
\bjournal{Appl. Environ. Microbiol.}
\bvolume{61}
\bpages{610--616}.
\bid{issn={0099-2240}, pmcid={1388350}, pmid={16534932}}
\end{barticle}
%
\bptok{imsref}%
% NOT OUTPUTED:
% issn = 0099-2240
% number = 2
% fjournal = Applied and environmental microbiology
\endbibitem

%b48 #&#
\bibitem[\protect\citeauthoryear{Shelef and Seiter}{2005}]{shelef2005}
%
\begin{bincollection}[author]
\bauthor{\bsnm{Shelef},~\bfnm{L.~A.}\binits{L.~A.}} \AND
\bauthor{\bsnm{Seiter},~\bfnm{J.}\binits{J.}}
(\byear{2005}).
\btitle{Indirect and miscellaneous antimicrobials}.
In \bbooktitle{Antimocrobials in Food},
\bedition{3rd} ed.
\bpages{573--598}.
\bpublisher{Taylor and Francis},
\blocation{Boca Raton, FL}.
\end{bincollection}
%
\bptok{imsref}%
\endbibitem

%b49 #&#
\bibitem[\protect\citeauthoryear{Skinner, Larkin and
Rhodehamel}{1994}]{skinner1994mathematical}
%
\begin{barticle}[author]
\bauthor{\bsnm{Skinner},~\bfnm{G.~U.~Y.~E.}\binits{G.~U.~Y.~E.}},
\bauthor{\bsnm{Larkin},~\bfnm{J.~W.}\binits{J.~W.}} \AND
\bauthor{\bsnm{Rhodehamel},~\bfnm{E.~J.}\binits{E.~J.}}
(\byear{1994}).
\btitle{Mathematical modeling of microbial growth: A review}.
\bjournal{J. Food. Saf.}
\bvolume{14}
\bpages{175--217}.
\end{barticle}
%
\bptok{imsref}%
\endbibitem

%b50 #&#
\bibitem[\protect\citeauthoryear{Spiegelhalter, Thomas and
Best}{1999}]{spiegelhalter1999winbugs}
%
\begin{bmisc}[author]
\bauthor{\bsnm{Spiegelhalter},~\bfnm{D.~J.}\binits{D.~J.}},
\bauthor{\bsnm{Thomas},~\bfnm{A.}\binits{A.}} \AND
\bauthor{\bsnm{Best},~\bfnm{N.~G.}\binits{N.~G.}}
(\byear{1999}).
\bhowpublished{WinBUGS version 1.2 user manual.
MRC Biostatistics Unit}.
\end{bmisc}
%
\bptok{imsref}%
\endbibitem

%b51 #&#
\bibitem[\protect\citeauthoryear{Spiegelhalter
et~al.}{2002}]{spiegelhalter2002bayesian}
%
\begin{barticle}[mr]
\bauthor{\bsnm{Spiegelhalter},~\bfnm{David~J.}\binits{D.~J.}},
\bauthor{\bsnm{Best},~\bfnm{Nicola~G.}\binits{N.~G.}},
\bauthor{\bsnm{Carlin},~\bfnm{Bradley~P.}\binits{B.~P.}} \AND
\bauthor{\bsnm{van~der Linde},~\bfnm{Angelika}\binits{A.}}
(\byear{2002}).
\btitle{Bayesian measures of model complexity and fit}.
\bjournal{J. R. Stat. Soc. Ser. B Stat. Methodol.}
\bvolume{64}
\bpages{583--639}.
\bid{doi={10.1111/1467-9868.00353}, issn={1369-7412}, mr={1979380}}
\end{barticle}
%
\bptok{imsref}%
% NOT OUTPUTED:
% issn = 1369-7412
% url = http://dx.doi.org/10.1111/1467-9868.00353
% number = 4
% fjournal = Journal of the Royal Statistical Society. Series B.
%Statistical Methodology
\endbibitem

%b52 #&#
\bibitem[\protect\citeauthoryear{Stern}{1996}]{stern1996neural}
%
\begin{barticle}[mr]
\bauthor{\bsnm{Stern},~\bfnm{Hal~S.}\binits{H.~S.}}
(\byear{1996}).
\btitle{Neural networks in applied statistics}.
\bjournal{Technometrics}
\bvolume{38}
\bpages{205--220}.
\bid{doi={10.2307/1270601}, issn={0040-1706}, mr={1411878}}
\bptnote{check related}%
\end{barticle}
%
\bptok{imsref}%
% NOT OUTPUTED:
% issn = 0040-1706
% url = http://dx.doi.org/10.2307/1270601
% number = 3
% coden = TCMTA2
% fjournal = Technometrics. A Journal of Statistics for the Physical,
%Chemical and Engineering Sciences
\endbibitem

%b53 #&#
\bibitem[\protect\citeauthoryear{Valero et~al.}{2007}]{valero2007searching}
%
\begin{barticle}[author]
\bauthor{\bsnm{Valero},~\bfnm{A.}\binits{A.}},
\bauthor{\bsnm{Hervas},~\bfnm{C.}\binits{C.}},
\bauthor{\bsnm{Garcia-Gimeno},~\bfnm{R.~M.}\binits{R.~M.}} \AND
\bauthor{\bsnm{Zurera},~\bfnm{G.}\binits{G.}}
(\byear{2007}).
\btitle{Searching for new mathematical growth model approaches for
\textit{Listeria monocytogenes}}.
\bjournal{J. Food Sci.}
\bvolume{72}
\bpages{M016--M025}.
\end{barticle}
%
\bptok{imsref}%
\endbibitem

%b54 #&#
\bibitem[\protect\citeauthoryear{Vehtari and
Lampinen}{2003}]{vehtari2003expected}
%
\begin{barticle}[auto]
\bauthor{\bsnm{Vehtari},~\bfnm{Aki}\binits{A.}} \AND
\bauthor{\bsnm{Lampinen},~\bfnm{Jouko}\binits{J.}}
(\byear{2003}).
\btitle{Expected utility estimation via cross-validation}.
\bjournal{Bayesian Stat.}
\bvolume{7}
\bpages{701--710}.
\end{barticle}
%
\bptok{imsref}%
\endbibitem

%b55 #&#
\bibitem[\protect\citeauthoryear{Watanabe}{2010}]{watanabe2010}
%
\begin{barticle}[author]
\bauthor{\bsnm{Watanabe},~\bfnm{S.}\binits{S.}}
(\byear{2010}).
\btitle{Equations of states in singular statistical estimation}.
\bjournal{Neural Netw.}
\bvolume{23}
\bpages{20--34}.
\end{barticle}
%
\bptok{imsref}%
\endbibitem

%b56 #&#
\bibitem[\protect\citeauthoryear{Wijtzes et~al.}{1995}]{wijtzes1995modelling}
%
\begin{barticle}[author]
\bauthor{\bsnm{Wijtzes},~\bfnm{T.}\binits{T.}},
\bauthor{\bsnm{De~Wit},~\bfnm{J.~C.}\binits{J.~C.}} \betal{et~al.}
(\byear{1995}).
\btitle{Modelling bacterial growth of \textit{Lactobacillus curvatus} as a
function of acidity and temperature}.
\bjournal{Appl. Environ. Microbiol.}
\bvolume{61}
\bpages{2533--2539}.
\end{barticle}
%
\bptok{imsref}%
\endbibitem

%b57 #&#
\bibitem[\protect\citeauthoryear{Wijtzes et~al.}{2001}]{wijtzes2001development}
%
\begin{barticle}[author]
\bauthor{\bsnm{Wijtzes},~\bfnm{T.}\binits{T.}},
\bauthor{\bsnm{Rombouts},~\bfnm{F.~M.}\binits{F.~M.}},
\bauthor{\bsnm{Kant-Muermans},~\bfnm{M.~L.~T.}\binits{M.~L.~T.}},
\bauthor{\bsnm{Van't~Riet},~\bfnm{K.}\binits{K.}} \AND
\bauthor{\bsnm{Zwietering},~\bfnm{M.~H.}\binits{M.~H.}}
(\byear{2001}).
\btitle{Development and validation of a combined temperature, water
activity, pH model for bacterial growth rate of \textit{Lactobacillus curvatus}}.
\bjournal{Int. J. Food Microbiol.}
\bvolume{63}
\bpages{57--64}.
\end{barticle}
%
\bptok{imsref}%
\endbibitem

%b58 #&#
\bibitem[\protect\citeauthoryear{Zwietering et~al.}{1990}]{zwietering1990modeling}
%
\begin{barticle}[author]
\bauthor{\bsnm{Zwietering},~\bfnm{M.~H.}\binits{M.~H.}},
\bauthor{\bsnm{Jongenburger},~\bfnm{I.}\binits{I.}},
\bauthor{\bsnm{Rombouts},~\bfnm{F.~M.}\binits{F.~M.}} \AND
\bauthor{\bsnm{Van't~Riet},~\bfnm{K.}\binits{K.}}
(\byear{1990}).
\btitle{Modeling of the bacterial growth curve}.
\bjournal{Appl. Environ. Microbiol.}
\bvolume{56}
\bpages{1875--1881}.
\end{barticle}
%
\bptok{imsref}%
\endbibitem

\end{thebibliography}
\end{document}